\documentclass[twoside,aps,superscriptaddress,pre,showpacs]{revtex4}

\usepackage{amsmath,amssymb,graphicx}

\newcommand{\Ha}{{\mathbf H}}

\newcommand{\e}{{\mathbf e}}
\newcommand{\om}{\omega}
\newcommand{\pa}{\partial}
\newcommand{\na}{\nabla}

\newcommand{\N}{{\mathbf N}}
\newcommand{\Nav}{\overline{\la N_z\ra}}

\newcommand{\m}{\boldsymbol{\mu}}
\newcommand{\te}{\theta}
\newcommand{\ph}{\varphi}
\newcommand{\ps}{\varphi_{\mathrm s}(t)}
\newcommand{\pus}{\varphi_{\mathrm{us}}(t)}
\newcommand{\la}{\langle}
\newcommand{\ra}{\rangle}
\newcommand{\phin}{\ph^\ast}

\newcommand{\Ord}{{\cal O}}
\newcommand{\Om}{\mathbf{\Omega}}
\newcommand{\ksi}{\boldsymbol{\xi}}
\newcommand{\eps}{\epsilon}
\begin{document} 

\title{Thermal ratchet effects in ferrofluids}

\author{Andreas Engel} 
\email[]{engel@theorie.physik.uni-oldenburg.de} 
\affiliation{Institut f\"ur Physik, Carl-von-Ossietzky-Universtit\"at,
     26111 Oldenburg, Germany, and \\
    CNRS-Laboratoire de Physique Th{\'e}orique, 
    3 rue de l'Universit\'e, 67000 Strasbourg, France.} 
\author{Peter Reimann} 
\affiliation{Theoretische Physik, Universit\"at Bielefeld, 
    33615 Bielefeld, Germany}
 
\date{\today} 
 
\begin{abstract} 
Rotational Brownian motion of colloidal magnetic particles in
ferrofluids under the influence of an oscillating external magnetic
field is investigated. It is shown that for a suitable time dependence
of the magnetic field, a noise induced rotation of the ferromagnetic 
particles due to rectification of thermal fluctuations takes
place. Via viscous coupling, the associated angular momentum is
transferred from the magnetic nano-particles to the carrier 
liquid and can then be measured as macroscopic torque on the fluid
sample. A thorough theoretical analysis of the effect 
in terms of symmetry considerations, analytical approximations, 
and numerical solutions
is given which is in accordance with recent experimental findings.
\end{abstract} 
 
\pacs{5.40.-a, 82.70.-y, 75.50.Mm}
 
\maketitle 
 

\section{Introduction}\label{intro}

Rotational Brownian motion of colloidal particles is a classical
subject in statistical physics \cite{Einstein}. Contrary to its 
translational cousin, {\it i.e.} ``usual'' Brownian motion, it does not
result in changes of the particle location and is hence not as easily
demonstrated in experiments. If, however, the suspended particles carry an
electric or magnetic moment their orientation couples to external
fields and the electric or magnetic relaxation properties of the
suspension are direct consequences of the rotational diffusion of
the constituting  particles. In particular, in the well developed theory of
ferrofluids the importance of rotational Brownian motion for the 
theoretical description of the often complex and surprising
hydrodynamic and magnetic properties was recognized already a long
time ago \cite{Sh}.  

Ferrofluids are suspensions of nano-size ferromagnetic grains in
carrier liquids like water or oil. They combine the hydrodynamic
properties of Newtonian fluids with the magnetic behaviour of
superparamagnets \cite{Ros}. Due to the viscosity of the carrier
liquid there is a coupling between the rotation of the magnetic
grains and the local vorticity of the hydrodynamic flow. This coupling
can, {\it e.g.},  be used to spin up ferromagnetic drops floating in a
non-magnetic fluid of the same density with the help of a 
{\em rotating magnetic field} \cite{BCP,LEM}. 
Complementary, a  rotational ferrofluid flow exposed to a static
magnetic field exhibits an enhanced shear viscosity
\cite{McTague}. Various related effects such as ``negative''
rotational viscosity \cite{ShMo}, magneto-vortical resonance
\cite{GHBCP}, and anomalously enhanced ac-response due to
coherent particle rotation \cite{HWM} have been investigated. They 
rely on the exchange of angular momentum between rotating 
particles and an oscillating magnetic field. In these cases, the 
imposed non-zero {\em flow vorticity} of the ferrofluid is crucial for 
breaking the symmetry between clockwise and counter-clockwise
particle rotation.

In the present paper we investigate a much more indirect and subtle
aspect of the interplay between rotational Brownian
motion of ferrofluid particles and their relaxation dynamics in an
external magnetic field. We will show that a suitably designed
time dependent external magnetic field {\em without net rotating
component} may rectify the fluctuations of the particle orientation and
set up a {\em noise-induced} rotation of the ferromagnetic grains. We
will hence investigate how angular momentum can be transferred from an 
{\em oscillating} magnetic field to a ferrofluid {\em at rest}. 
The effect was predicted theoretically and demonstrated experimentally
in a previous short communication \cite{EMRJ}. In the present paper we
provide much more details on the theoretical description and add
several new results. 

The extraction of directed motion from random fluctuations is an old and 
controversial problem in statistical mechanics with a long and
interesting history \cite{Maxwell,Smo,Fey,HaRe}. Although excluded by
the second law of thermodynamics for equilibrium systems, rectification
of fluctuations {\em is} 
possible in systems driven sufficiently far away from thermal
equilibrium \cite{Reirev,Ast}. The problem has gained renewed
attention under the trademarks of ``thermal ratchets'' and ``Brownian motors''
due to its possible relevance for biological
transport \cite{Mag,JAP} and the prospects of nano-technology 
\cite{RSAP,OuBo,Lin}. 

Ferrofluids are ideal systems to investigate such fluctuation driven
transport phenomena and also to demonstrate them experimentally \cite{EMRJ}: As already
discussed above, the rotational dynamics of the ferromagnetic 
grains is strongly influenced by thermal fluctuations. Appropriate
time-dependent potentials can be easily designed with the 
help of external magnetic fields. Finally, directed rotational
transport in ferrofluids should manifest itself as systematic 
rotation of the ferromagnetic nano-particles. This in turn can be easily
detected from the resulting {\it macroscopic} torque on the carrier liquid. 
Various somewhat related but still quite different phenomena in rotational 
dynamical systems are treated in \cite{tur89,PaWe,fla02,CKW,rit98}. 

Throughout the paper we will use two basic approximations which
simplify the analysis considerably and which are rather common for
ferrofluids \cite{Ros,Sh}. The first is to neglect Neel-relaxation of
the magnetization, {\it i.e.} the rotation of the magnetization vector with
respect to the ferromagnetic particle. This is justified for particle
sizes that are not too small and amounts to assuming that the
magnetic moments are firmly attached to the geometry of the
particles. Any reorientation of the magnetic moment hence requires a
rotation of the particle as a whole. The second approximation is to
neglect dipole-dipole interactions between the particles. Although these
interactions may be important in concentrated ferrofluids, they are
negligible in sufficiently diluted ones. The ratchet mechanism of
central interest in the present investigation can operate without any
particle-particle interaction. We will hence assume that our
ferrofluids are sufficiently diluted and use a  
{\em single particle} model. For a more quantitative assessment 
of the role of the dipole-dipole interactions see {\it e.g.}
\cite{WHM} and references therein. 

The paper is organized as follows. In section \ref{sec:model} we
introduce the general framework for describing the rotational
Brownian motion of a single suspended ferromagnetic particle in an
external magnetic field. Section \ref{sec:symm} gives a detailed
account of the symmetries characteristic of our system and discusses
how these symmetries influence the possibility for a ratchet mechanism
to operate. In section \ref{sec:FPE} the central Fokker-Planck
equation describing our system is analyzed. Besides an investigation
of the weak noise limit we describe the numerical solution of the
Fokker-Planck equation and compare the results with two different 
approximate treatments yielding  analytical expressions for the
transferred angular momentum. Finally, section \ref{sec:conc} contains
some conclusions.


\section{General framework}\label{sec:model}

We consider the rotational motion of a spherical particle of volume
$V$ and magnetic moment $\m$ immersed in a liquid with dynamic viscosity 
$\eta$ and subject to a horizontal, time dependent, spatially homogeneous
magnetic field $\Ha$. The field is composed of a constant part $H_x$
parallel to the $x$-axis and an oscillatory part $H_y(t)$ with period
$2\pi/\om$ along the $y$-direction  
\begin{equation}\label{deffield}
  \Ha = (H_x,H_y(t),0) \quad,\quad 
    H_y(t+2\pi/\om)=H_y(t) \ .
\end{equation}
Different choices for the time dependence of $H_y$ are of interest. In
the present paper we will mainly discuss two cases which are each 
representative for a whole class. Our first standard choice is 
\begin{equation}\label{foft1}
  H_y(t)= H_y^{(1)}\cos(\om t)+ H_y^{(2)}\sin(2\om t+\delta) \ ,
\end{equation}
where the amplitudes $H_y^{(1,2)}$, and the phase $\delta$ are control
parameters. The main features of this time dependence are a zero
average over one period and the presence of a higher harmonic of the
basic frequency. As a second example, we will also discuss the form 
\begin{equation}\label{foft2}
  H_y(t)= H_y^{(0)} + H_y^{(1)} \cos(\om t)\ ,
\end{equation}
for which the average over one period is different from zero. It
arises naturally if the constant field component is not perpendicular
to the time dependent one. In any case, the magnetic field is of a pure
oscillatory character, {\it i.e.} it 
{\em does not contain a net rotating component}.  

The orientation of the particle at time $t$ is described by the unit
vector $\e(t)=\m(t)/\mu$ where $\mu$ denotes the modulus of the
magnetic moment. The time evolution of $\e$ is given by 
\begin{equation}\label{evol1}
  \pa_t \e= \Om\times\e \ ,
\end{equation}
where $\Om(t)$ denotes the instantaneous angular velocity of the
particle. Changes of $\Om$ are due to torques on the
particle. Denoting by $\nabla$ the angular part of the three dimensional 
Nabla operator, the magnetic torque 
\begin{equation}\label{defmagtor}
 \N_{\mathrm{mag}}=-\e\times\nabla U = \mu\; \e\times\Ha
\end{equation}
derives from the potential energy 
\begin{equation}\label{defU}
  U(\e,t)=-\mu\; \e \cdot \Ha(t)
\end{equation}
of a magnetic dipole in an external field \cite{LLVIII}. 
Further, the viscosity $\eta$ of the carrier liquid gives rise to
a viscous torque \cite{LLVI}
\begin{equation}\label{defvisctor}
  \N_{\mathrm{visc}}=-6\eta V \Om \ .
\end{equation}
Additionally, the interaction between the rotating particle and the
surrounding liquid also causes thermal fluctuations which
generate a stochastic torque \cite{CKW}
\begin{equation}\label{defstochtor}
  \N_{\mathrm{stoch}}=\sqrt{12\eta V k_B T}\;\ksi(t) \ .
\end{equation}
Here, $\ksi(t)$ is a vector of independent,
$\delta$-correlated Gaussian noise
sources of zero mean, the noise intensity 
is related to the temperature $T$ and the dissipation (\ref{defvisctor}) 
of the carrier liquid by the fluctuation-dissipation relation, and $k_B$ stands
for Boltzmann's constant. Denoting the moment of inertia of the particle by
$I$, the equation of motion for $\Om$ acquires the form 
\begin{equation}\label{evol2}
  I \pa_t \Om + 6 \eta V \Om = \mu\ \e\times\Ha + \sqrt{12\eta V k_B T}\;\ksi(t) \ .
\end{equation}
Eqs.(\ref{evol1}) and (\ref{evol2}) form a closed set of
equations for the description of the rotational motion of the
particle. Using experimentally relevant parameter values (density of
the particle $\rho\simeq 4\cdot10^{3}$ kg/m$^3$, particle radius
$R\simeq 10$nm, viscosity $\eta\simeq 10^{-1}\dots 10^{-3}$ Pas, time scale
$\Om/\pa_t \Om\simeq 10^{-4}$ s) 
we find that the first term on the l.h.s. of eq.~(\ref{evol2}) is five
to seven orders of magnitude smaller than the second one. We may
hence safely neglect inertial effects \cite{Sh} and find for $\Om$ in the
overdamped limit
\begin{equation}\label{eqOm2}
  \Om = \frac{\mu}{6 \eta V}\; \e\times\Ha + 
             \sqrt{2D}\;\ksi(t) \ ,
\end{equation}
where we have introduced $D=k_BT/6\eta V$.
Using this result in (\ref{evol1}) yields a closed equation for the
time evolution of $\e$ 
\begin{equation}\label{evol3}
  \pa_t \e= \frac{\mu}{6 \eta V}\; (\e\times\Ha)\times\e +
                \sqrt{2D}\;\ksi(t)\times \e \ .      
\end{equation}
It is convenient to introduce dimensionless units. To this end we
measure time in units of the inverse of the external driving
frequency, $t\mapsto t/\om$, use $6\eta V \om/\mu$ as unit for the
magnetic field strength, $\Ha\mapsto (6\eta V \om/\mu)\;\Ha$, and
rescale the noise intensity according to 
$D \mapsto \om\; D$. The evolution equation for the 
orientation $\e$ then reads 
\begin{equation}\label{evol4}
  \pa_t \e= (\e\times\Ha)\times\e + \sqrt{2D}\;\ksi(t)\times \e \ . 
\end{equation}
Introducing the Brownian relaxation time 
\begin{equation}\label{deftauB}
  \tau_B=\frac{3\eta V}{k_B T} \ ,
\end{equation}
we note that the rescaled, dimensionless noise intensity $D$ occurring in
(\ref{evol4}) just gives the ratio between the relevant deterministic
and stochastic time scales in the system:
\begin{equation}\label{defD}
  D=\frac{k_B T}{6 \eta V \om}=\frac{1}{2\tau_B\; \om} \ .
\end{equation}

To proceed, we parametrize the orientation $\e$ of the magnetic
particle by two angles $\te$ and $\ph$ according to 
\begin{equation}\label{defe}
  \e =(\sin\theta\cos\ph,\sin\theta\sin\ph,\cos\theta) \ .
\end{equation}
{F}rom (\ref{evol4}) we then find the following Langevin equations for
the time evolution of these angles \cite{RaEn}
\begin{align}\label{lan1}
  \pa_t \te &=-\pa_\te U + D \cot\te + \sqrt{2D}\;\xi_\te(t)\\\label{lan2}
  \pa_t \ph &=-\frac 1 {\sin^2\te} \pa_\ph U 
                +\frac{\sqrt{2D}}{\sin\te} \;\xi_\ph(t) \ .
\end{align}
In the dimensionless units adopted, the noise intensity $D$ is given by
(\ref{defD}) and the potential (\ref{defU}) takes the form 
\begin{equation}\label{defpot}
 U(\theta,\ph,t)= - \sin\te\; (H_x \cos\ph+ H_y(t)\sin\ph)\ .
\end{equation}
The thermal fluctuations $\xi_\te(t)$ and $\xi_\ph(t)$ are given by two
independent, $\delta$-correlated Gaussian noise sources of zero mean. 
Note that in (\ref{lan2}) we are dealing with
multiplicative white noise but since the  multiplicative function 
is independent of $\ph(t)$, no ambiguity (Ito-
vs. Stratonovich-interpretation \cite{ris82}) arises. 

It is instructive also to consider the simplified situation of a
one-dimensional dynamics in which the vector $\e$ is assumed to be
constrained to the horizontal $x$-$y$-plane. Accordingly, (\ref{lan1})
is replaced by $\te\equiv \pi/2$ and (\ref{lan2}) simplifies to 
\begin{equation}\label{lan3}
\pa_t \ph=F(\ph ,t) + \sqrt{2 D}\,\xi_\ph(t)
\end{equation}
where 
\begin{equation}\label{defF}
  F(\ph,t)=-H_x\,\sin\ph + H_y(t)\,\cos\ph \ .
\end{equation}

The observable of foremost interest in the present investigation is 
the time and ensemble averaged torque (\ref{defmagtor}) exerted by the magnetic field
upon the magnetic particle in the long time limit, {\it i.e.} after
initial transient have died out. Since the magnetic field (\ref{deffield}) is
constrained to the $x$-$y$ plane, only the $z$-component of this
magnetic torque can be different from zero. Suppressing the subscript
$_\mathrm{mag}$ in (\ref{defmagtor}) from now on, we denote the averaged
$z$-component of the magnetic torque by $\Nav$, where 
$\langle\dots\rangle$ stands for the ensemble average over the 
different realizations of the noise terms in
(\ref{lan1}), (\ref{lan2}) and the overbar represents the time
average over one period of the magnetic field. Using (\ref{deffield})
and (\ref{defmagtor}) we get 
\begin{equation}\label{torav}
  \Nav=\frac 1 {2\pi}\int\limits_0^{2\pi} dt\,
  \langle \sin\te(t)\, (-\,H_x\,\sin\ph(t)+H_y(t)\,\cos\ph(t)) \rangle
  \ .
\end{equation}
Exploiting (\ref{defpot}) and (\ref{lan2}) one readily finds the
equivalent expressions 
\begin{equation}\label{torav'}
  \Nav=\frac 1 {2\pi}\int\limits_0^{2\pi} dt\,
  \langle  \pa_\ph U(\te (t),\ph (t),t)\rangle 
      =\frac 1 {2\pi}\int\limits_0^{2\pi} dt\,
  \langle \pa_t \ph(t)\,\sin^2\theta(t) \rangle
\end{equation}
For reasons of ergodicity, the ensemble average in (\ref{torav}) is
equivalent to a time average of a single realization over an infinite
time interval. Then the extra time average over one period of the
external driving drops out and we are left with 
\begin{equation}\label{torav''}
  \Nav=\lim_{(t_f-t_i)\to\infty}\frac 1 {t_f-t_i}\int\limits_{t_i}^{t_f} dt\,
   \pa_t \ph(t)\,\sin^2\te(t)
\end{equation}
and similarly for the equivalent expressions in (\ref{torav}),
(\ref{torav'}). 

\begin{figure} 
  \includegraphics[width=0.5\columnwidth]{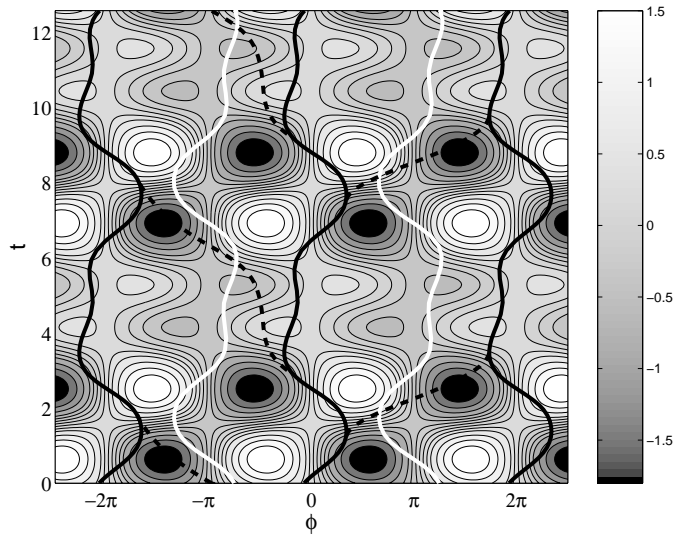}  
  \caption{\label{fig:fig1} Space-time plot of the potential
$U(\theta=\pi/2,\ph,t)$ from eq.(\ref{defpot}) for the time dependence
(\ref{foft1}) with 
$H_x=0.3,\, H_y^{(1)}=H_y^{(2)}=1$, and $\delta=0$. Dark and bright
regions correspond to small and large values of $U$, respectively. 
In the long-time limit, the deterministic dynamics (\ref{lan1}),
(\ref{lan2}) with $D=0$ approaches $\theta(t) =\pi/2$ and a 
periodical $\ph (t)$, oscillating back and forth as 
represented by either of the full black lines. In the presence of a
small amount of noise, occasional transitions across the unstable deterministic
orbits shown as full white lines become possible which are
schematically indicated by the dashed lines. The spatial asymmetry and
temporal anharmonicity of the potential conspire to yield slightly
different rates for noise induced increments and decrements of $\ph $,
respectively. As a result a noise driven rotation of the particles
arises. For a detailed discussion see section \ref{inst1d}.}
\end{figure} 

In the absence of thermal fluctuations, the particle orientation
is governed by the overdamped deterministic relaxation dynamics given
by (\ref{lan1}), (\ref{lan2}) with $D=0$. Hence, at any given moment,
the orientation $\e(t)$ tends to align with the instantaneous magnetic
field $\Ha(t)$ which lies in the $x$-$y$-plane. One can then show
\cite{BeEn} that $\te(t)$ converges to $\pi/2$ for $t\to\infty$ and that
$\ph(t)$ approaches a {\it periodic} long time behaviour. This implies
via (\ref{torav''}) that  
\begin{equation}\label{eq24}
  \Nav=\lim_{(t_f-t_i)\to\infty}\frac {\ph(t_f)-\ph(t_i)}{t_f-t_i}=0\,
\end{equation}
{\it i.e.}, in the absence of thermal fluctuations no particle
rotation will occur  and no average torque can arise. An explicit
example is displayed in Fig.~\ref{fig:fig1}.  

This scenario changes fundamentally in the presence of fluctuations,
{\it i.e.} if $D\neq 0$. As shown qualitatively in Fig.~\ref{fig:fig1}, for
small noise intensities, the time dependence of $\ph(t)$ and $\te(t)$
will still closely follow the deterministic trajectories, except that now
also rare, fluctuation induced {\em transitions} between different
deterministic solutions may occur. If the rates of forward (increasing
$\ph$ by $2\pi$) and backward transitions (decreasing $\ph$ by $2\pi$)
are different from each other, then on the average a net rotation of the particle
will occur, implying with (\ref{eq24}) that
$\Nav \neq 0$. This is a manifestation of the
so-called ratchet effect \cite{Reirev} in which an unbiased potential and undirected
Brownian fluctuations cooperate to produce directed transport. The
detailed operation of the ratchet mechanism for the rotational
motion of a colloidal particle is the main focus of the present
paper. 


\section{Symmetries}\label{sec:symm}

Symmetry considerations admit conclusions from the
equations of motion without actually solving them. For the
investigation of ratchet systems, studies of symmetry turned out to be
valuable since directed transport generically occurs if it is not forbidden
by symmetries, a statement referred to as Curie's principle
\cite{Reirev,curie,ajd94,kan99,wei00,fla00,cil01,neu01,yev01,yan01,rei01c,den02}.
Here we specifically address the question which
transformations of the magnetic field components
$H_x$ and $H_y(t)$ in (\ref{deffield}) leave the
average $z$-component of the torque invariant, $\Nav\mapsto \Nav$, and
which lead to its inversion, $\Nav\mapsto -\Nav$. 

{F}rom Eq. (\ref{torav''}) one readily concludes that the following 
transformations of $\ph$ and $\te$ imply the indicated transformation
behavior of $\Nav$:
\begin{alignat}{5}\label{1a}
&\ph(t)\mapsto\ph(t)+\Delta\ph\qquad  &\text{and}\qquad& 
   \te(t)\mapsto\te(t) \qquad& \text{implies}\qquad \Nav&\mapsto\Nav\\
\label{1b}
&\ph(t)\mapsto-\ph(t)+\Delta\ph\qquad  &\text{and}\qquad& 
   \te(t)\mapsto\te(t) \qquad& \text{implies}\qquad \Nav&\mapsto-\Nav\\
\label{1c}
&\ph(t)\mapsto\ph(t)\qquad  &\text{and}\qquad& 
   \te(t)\mapsto-\te(t)+\pi \qquad& \text{implies}\qquad \Nav&\mapsto\Nav\\
\label{1d}
&\ph(t)\mapsto\ph(t+\Delta t)\qquad  & \text{and}\qquad& 
   \te(t)\mapsto\te(t+\Delta t) \qquad& \text{implies}\qquad \Nav&\mapsto\Nav\\
\label{1e}
&\ph(t)\mapsto\ph(-t)\qquad  &\text{and}\qquad& 
   \te(t)\mapsto\te(-t) \qquad& \text{implies}\qquad \Nav&\mapsto-\Nav
\end{alignat}
where $\Delta\ph$ and $\Delta t$ are arbitrary. Observing that the
values of both the original and the transformed $\te(t)$ are restricted
to the interval $[0,\pi]$ and that the entire interval will indeed be
sampled by some realization of the dynamics (\ref{lan1}), (\ref{lan2}),
it follows that no transformations of the form 
$\te(t)\mapsto\te(t)+\Delta\te$ with $\Delta\theta\neq 0$
or $\te(t)\mapsto-\te(t)+\Delta\te$ with $\Delta\theta\neq \pi$
are possible. Finally, in order to conclude $\Nav\mapsto -\Nav$ in
(\ref{1e}) one has to exploit that in (\ref{torav''}) the limit
$(t_f-t_i)\to\infty$ is not necessarily tantamount to $t_f\to\infty$
with $t_i$ kept fixed, but can also be realized by letting $t_i\to
-\infty$ with $t_f$ kept fixed. 

Closer inspection of (\ref{torav''}) reveals that any further
symmetry transformation that leaves $\Nav$ invariant or changes its
sign can be composed of the ``elementary'' transformations
(\ref{1a})-(\ref{1e}). 

Next we ask ourselves what equations of motion govern the transformed 
$\theta(t)$, $\ph(t)$ defined by (\ref{1a})-(\ref{1e}). 
In the simplest case (\ref{1a}) one readily concludes from (\ref{lan1})
and (\ref{lan2}) that the transformed $\theta(t)$, $\ph(t)$ satisfy
the same equations of motion (\ref{lan1}) and (\ref{lan2}) but with a
transformed potential 
$U(\theta,\ph,t)\mapsto U(\theta,\ph-\Delta\ph,t)+\Delta U$ with
arbitrary $\Delta U$. Turning to the second transformation (\ref{1b})
one finds similarly that the transformed $\theta(t)$ and $\ph(t)$
satisfy the equations of motion (\ref{lan1}) and (\ref{lan2}) but now
with a transformed potential 
$U(\theta,\ph,t)\mapsto U(\theta,-\ph+\Delta,t)+\Delta U$ 
and in addition with a transformed noise 
$\xi_\ph(t)\mapsto-\xi_\ph(t)$. Since the statistical properties of
the two noises $-\xi_\ph(t)$ and $\xi_\ph(t)$ are equal, 
{\it i.e.} they are identical stochastic processes, the same follows for
the corresponding processes given by (\ref{lan1}) and (\ref{lan2}).
In other words, to each realization of the original $\theta(t)$,
$\ph(t)$ corresponds a realization of the transformed $\theta(t)$,
$\ph(t)$ occurring with the same probability, and vice versa. Since
the average torque in (\ref{torav''}) 
is independent of the specific realization $\theta(t)$, $\ph(t)$
with probability 1 ({\it i.e.} up to a set of realizations of measure zero)
we can conclude that in this case $\Nav\mapsto -\Nav$. Similar
considerations can be done for the cases given by (\ref{1c}) and
(\ref{1d}). The last transformation (\ref{1e}) is insofar special as a
mapping between the dynamics of the original and the transformed
variables is possible only, if the term $D\cot\theta (t)$ in
(\ref{lan1}) identically vanishes, {\it i.e.} if $\theta (t)\equiv
\pi/2$. We therefore find the following implications of
transformations in the potential $U(\te,\ph,t)$:
\begin{alignat}{4}\label{2a}
&U(\te,\ph,t)\mapsto U(\te,\ph-\Delta\ph,t)+\Delta U\qquad&
  \text{implies}\qquad \Nav&\mapsto\Nav\\
\label{2b}
&U(\te,\ph,t)\mapsto U(\te,-\ph+\Delta\ph,t)+\Delta U\qquad&
  \text{implies}\qquad \Nav&\mapsto-\Nav\\
\label{2c}
&U(\te,\ph,t)\mapsto U(-\te+\pi,\ph,t)+\Delta U\qquad&
  \text{implies}\qquad \Nav&\mapsto\Nav\\
\label{2d}
&U(\te,\ph,t)\mapsto U(\te,\ph,t+\Delta t)+\Delta U\qquad&
  \text{implies}\qquad \Nav&\mapsto\Nav\\
\label{2e}
&U(\te,\ph,t)\mapsto -U(\te,\ph,-t)+\Delta U\qquad&
  \text{implies}\qquad \Nav&\mapsto-\Nav 
                \qquad(\text{provided}\;\; \te(t)\equiv \pi/2) \ ,
\end{alignat}
where $\Delta U$ is an arbitrary constant. 

Finally, we have to determine all those transformations of 
the magnetic field components $H_x$ and $H_y(t)$ in (\ref{deffield}) 
which correspond via (\ref{defpot}) to transformations of the
potential $U(\theta,\ph,t)$ given in (\ref{2a})-(\ref{2e}).
Since changes in $H_x$ and $H_y(t)$ always imply changes in the
potential $U$ that depend on $\te$ and $\ph$, we can only induce
transformations with $\Delta U=0$. The only non-trivial way to realize
(\ref{2a}) is $(H_x,\, H_y(t)) \mapsto (-H_x,\, -H_y(t))$ combined
with $\Delta\ph = (2n+1)\pi$ with an arbitrary integer $n$. 
Similarly, (\ref{2b}) and (\ref{2d}) are equivalent to 
$(H_x,\, H_y(t)) \mapsto (H_x,\, -H_y(t))$ and 
$(H_x,\, H_y(t)) \mapsto (H_x,\, H_y(t+\Delta t))$. 
On the other hand (\ref{2c}) only admits the trivial realization 
$(H_x,\, H_y(t)) \mapsto (H_x,\, H_y(t))$. 
Finally, (\ref{2e}) can be implemented by 
$(H_x,\, H_y(t)) \mapsto (-H_x,\, -H_y(-t))$. In summary, we find that
\begin{alignat}{4}\label{3a}
&(H_x,\, H_y(t)) \mapsto (-H_x,\, -H_y(t))\qquad&
  \text{implies}\qquad \Nav&\mapsto\Nav\\
\label{3b}
&(H_x,\, H_y(t)) \mapsto (H_x,\, -H_y(t))\qquad&
  \text{implies}\qquad \Nav&\mapsto-\Nav\\
\label{3c}
&(H_x,\, H_y(t)) \mapsto (H_x,\, H_y(t))\qquad&
  \text{implies}\qquad \Nav&\mapsto\Nav\\
\label{3d}
&(H_x,\, H_y(t)) \mapsto (H_x,\, H_y(t+\Delta t))\qquad&
  \text{implies}\qquad \Nav&\mapsto\Nav\\
\label{3e}
&(H_x,\, H_y(t)) \mapsto (-H_x,\, -H_y(-t))\qquad&
  \text{implies}\qquad \Nav&\mapsto-\Nav 
                \qquad(\text{provided}\;\; \te(t)\equiv \pi/2) \ ,
\end{alignat}
where the trivial result (\ref{3c}) is listed only for the sake of
completeness. 

While (\ref{3a})-(\ref{3d}) are intuitively more or less obvious,
(\ref{3e}) respectively (\ref{2e}) is not. The additional 
condition $\theta(t)\equiv \pi/2$ shows that the constrained dynamics
(\ref{lan3}), (\ref{defF}) has an extra symmetry which is lost if the
full two-dimensional dynamics (\ref{lan1}), (\ref{lan2}) is
considered. 

We note that (\ref{3a})-(\ref{3e}) could also have been obtained
in a more direct way. Our somewhat more involved line of reasoning has
the advantage that we can exclude that there are any transformations
other than (\ref{3a})-(\ref{3e}) and combinations thereof
which would leave $\Nav$ invariant or change its sign.

In order to see that the above considerations already allow some
non-trivial conclusions about the possibility of angular momentum
transfer from the oscillating magnetic field to the particle we
combine (\ref{3a}) and (\ref{3b}), (\ref{3b}) and (\ref{3d}), and
(\ref{3a}) and (\ref{3e}) to obtain  
\begin{alignat}{4}\label{3f}
&(H_x,\, H_y(t)) \mapsto (-H_x,\, H_y(t))\qquad&
  \text{implies}\qquad \Nav&\mapsto-\Nav \\
\label{3g}
&(H_x,\, H_y(t)) \mapsto (H_x,\, -H_y(t+\Delta t))\qquad&
  \text{implies}\qquad \Nav&\mapsto-\Nav \\
\label{3h}
&(H_x,\, H_y(t)) \mapsto (H_x,\, H_y(-t))\qquad&
  \text{implies}\qquad \Nav&\mapsto-\Nav 
                \qquad(\text{provided}\;\; \te(t)\equiv \pi/2)
\end{alignat}
respectively. From (\ref{3f}) we immediately can infer that the constant
field in $x$-direction is indispensable for a non-zero average torque,
\begin{equation}\label{4a}
H_x = 0 \ \ \Rightarrow \ \ \Nav = 0 \ .
\end{equation}
Similarly (\ref{3g}) implies that for a time dependence obeying 
$H_y(t) = -H_y(t+\Delta t)$ for some $\Delta t$ the average torque has
to vanish as well 
\begin{equation}\label{4b}
H_y(t) = -H_y(t+\Delta t) \ \ \Rightarrow \ \ \Nav = 0 \ .
\end{equation}
Choosing in particular $\Delta t=\pi$, {\it i.e.} half the period of the
external driving, we realize that an oscillating field $H_y(t)$ with a
Fourier expansion containing only odd multiples of the basic frequency 
\begin{equation}\label{4c}
H_y(t) = \sum_{n=1,3,5,...} H_y^{(n)}\,\cos (n t+\delta_n)
\end{equation}
will result in a zero average torque $\Nav$, irrespective of
the particular choices of the amplitudes $H_y^{(n)}$ and phases
$\delta_n$. Eqs.(\ref{4a}) and (\ref{4c}) motivate our choices
(\ref{foft1}) and (\ref{foft2}) as simple time dependencies of the
oscillating field resulting in a non-zero average torque $\Nav$. 

For the special time dependence (\ref{foft1}) we find from (\ref{3g})
\begin{equation}\label{Nofdel}
  \Nav(\delta+\pi)=-\Nav(\delta)
\end{equation}
As a consequence, upon continuously varying $\delta$ we can infer that
there must exist a $\delta_0\in[0,\pi)$ such that 
$\Nav =0$ for $\delta=\delta_0$ or $\delta=\delta_0+\pi$. It is important to note that this
case of zero averaged torque is qualitatively different from the
situations described by eqs. (\ref{4a}) and (\ref{4b}) since it is due
to the {\em fine tuning of a parameter} rather than resulting from an
underlying symmetry. Consequently, upon variation of $\delta$ around
$\delta_0$ the average torque $\Nav$ changes sign, which is a
realization of a so-called {\em current inversion} \cite{Reirev}. 
Furthermore, by fixing $\delta=\delta_0$, a sign change of $\Nav$ will generically
also occur upon variation of any other parameter of the system. 
Interesting examples for such a parameter include the driving
frequency $\omega$, the noise intensity $D$, and the
particle volume $V$. In the latter case we hence face the intriguing
possibility that in a polydisperse ferrofluid under the same
experimental conditions the larger and smaller particles will rotate
in {\em opposite} directions. 

We finally note that in the case where the dynamics is constrained to
the $x$-$y$-axis, the additional symmetry (\ref{3h}) implies that 
$\delta_0=\pi/2$. Consequently the $\delta$-values where
the torque changes sign are fixed by the additional symmetry and do not
depend on the other parameters of the problem. Hence no current
reversal upon changes in these parameters is possible. This is the
main difference between the  full two-dimensional dynamics
(\ref{lan1}), (\ref{lan2}) and its simplified one-dimensional
counterpart (\ref{lan3}), see also \cite{SeMa,ChMi,GoHa,fla00}.


\section{Analysis of the Fokker-Planck Equation}\label{sec:FPE}

A quantitative analysis of the overdamped Brownian rotation of a single
ferrofluid particle in a time dependent external field can be built
on the numerical simulation of the Langevin equations (\ref{lan1}),
(\ref{lan2}). Such simulations unambiguously show the occurrence of
noise induced rotation of the particle in an oscillating field
\cite{RaEn}. For a detailed study of the influence of the various
parameters of the problem on the transferred torque $\Nav$ it is, 
however, more convenient to start with the equivalent Fokker-Planck
Equation (FPE) for the probability density $P=P(\te,\ph,t)$ for the
orientation (\ref{defe}) of the particle 
\begin{align}\nonumber
  \pa_t P=& \na (P \, \na U)+D \na^2 P \\
   =&\frac{1}{\sin\te}\, \pa_\te\left(\sin\te \,P \,\pa_\te U\right)
    +\frac{1}{\sin^2\te}\,\pa_\ph \left(P \,\pa_\ph U\right)
    +\frac{D}{\sin\te}\, \pa_\te\left(\sin\te \, \pa_\te P\right)
    +\frac{D}{\sin^2\te}\, \pa^2_\ph P \ .\label{FPE}
\end{align}
Here $\na$ denotes the angular part of the three dimensional Nabla
operator. The equivalence of (\ref{lan1}), (\ref{lan2}) and
(\ref{FPE}) is, {\it e.g.}, shown in \cite{RaEn}. For a
time-periodic potential $U(\te,\ph,t)$, the solution
$P(\te,\ph,t)$ of (\ref{FPE}) will also be periodic after initial
transients have died out. From this solution we can determine the average
orientation of the particle
\begin{equation}
  \langle \e(t) \rangle=\int_0^{2\pi}d\ph \int_0^\pi d\te \sin\te
     (\sin\te \cos\ph, \sin\te \sin\ph, \cos\te)\,P(\te,\ph,t)
   \label{eq48}
\end{equation}
with the help of which the average torque from
(\ref{torav})-(\ref{torav'}) can be calculated according to
(\ref{defmagtor}) in yet another way, namely
\begin{equation}\label{torfpe}
  \Nav=\frac{1}{2\pi}\int_0^{2\pi}\!\! dt\; [\langle \e(t) \rangle \times
  \Ha(t) ]_z .
\end{equation}

In the present section we discuss the numerical and approximate
analytical solution of the Fokker-Planck equation (\ref{FPE}) and the
corresponding results (\ref{eq48}), (\ref{torfpe})
for the averaged torque $\Nav$. To first
substantiate the intuitive understanding of the ratchet effect in our
system advocated with Fig.~\ref{fig:fig1}, we start by considering the
weak noise limit $D\to 0$. To keep the analysis simple we will
restrict ourselves in this part to the simplified one-dimensional
model defined by (\ref{lan3}), (\ref{defF}). In the second subsection
we detail the numerical methods used to solve the FPE for the general
case. Finally we present some analytical results from the perturbative
solution of the FPE.


\subsection{Instantons in $d=1$}\label{inst1d}
Our starting point is the simplified one-dimensional model 
dynamics (\ref{lan3}), (\ref{defF}) with a $\delta$-correlated,
unbiased Gaussian noise source $\xi_\ph(t)$.
The deterministic dynamics described by (\ref{lan3}) with $D=0$ has 
a family of stable and unstable periodic orbits 
$\ps\pm 2\pi n$ and $\pus\pm 2\pi n$, respectively, where $n$
is an arbitrary integer (cf. Fig.~\ref{fig:fig1}) \cite{BeEn}. Hence 
deterministically the particle simply follows the direction of the
magnetic field with a certain time lag due to viscous damping and no 
particle rotation can occur. 

For any non-zero noise intensity $D>0$, there will be few
stochastic transitions between the deterministic trajectories and
their detailed form in the limit $D\to 0$ can be
determined. Comparing the associated rates 
for forward (increasing $\ph$ by $2\pi$) and backward (decreasing
$\ph$ by $2\pi$) phase slips we may directly obtain direction and
magnitude of the noise-driven transport by means of (\ref{eq24}).

The analysis of this section builds on the path integral
representation for the transition probability of a Markovian
stochastic process \cite{FeHi,Schu,ChDe}. The main
complication is the {\em time dependence} of the potential
$U(\te,\ph,t)$. In notation and general strategy we follow the recent
detailed analysis of noise-driven escape over oscillating barriers
\cite{LRH}. We will only determine the dominating exponential term in
the transition probabilities. A more detailed analysis including also
the prefactor can be done along the lines of ref.~\cite{LRH}.

The transition probability 
$p(\ph_f,t_f|\ph_i,t_i)$ of the Markov process described by
(\ref{lan3}) can be written in the form
\begin{equation}\label{pathint}
  p(\ph_f,t_f|\ph_i,t_i)=
     \int\limits_{\ph(t_i)=\ph_i}^{\ph(t_f)=\ph_f}
    {\cal D}\ph(\cdot)\exp(-\frac{S[\ph(\cdot)]}{D}) \ ,
\end{equation}
where the action functional $S[\ph (\cdot)]$ is given by
\begin{equation}\label{action}
  S[\ph(\cdot)]=\frac 1 4 \int_{t_i}^{t_f} dt 
      \big(\pa_t\ph(t)-F(\ph(t),t)\big)^2.
\end{equation}
with $F(\ph,t)$ defined in (\ref{defF}). 
The intuitive understanding behind this representation is that {\em all}
possible trajectories $\ph(t)$ starting at initial time $t_i$ in
$\ph_i$ and arriving at final time $t_f$ in $\ph_f$
contribute to the transition probability $p(\ph_f,t_f|\ph_i,t_i)$,
each with a weight related to the value of the action $S$ evaluated
along this trajectory. In the weak noise limit $D\to 0$ a Laplace
argument tells us that only those transitions $\phin(t)$ contribute
significantly to $p(\ph_f,t_f|\ph_i,t_i)$ for which the action is
minimal. All others have probabilities which are exponentially small
for $D\to 0$. In the weak noise limit we are hence able to determine
the precise form of the (dominating) stochastic transitions by solving
a variational problem.  

Under rather general conditions which are satisfied in our particular
setting one can show that the transition rate between two stable
orbits of the dynamics is determined by the probability to reach from
the starting stable orbit the unstable orbit separating the two stable
ones \cite{LRH}. The subsequent relaxation to the final stable orbit is
then purely deterministic and does not contribute to the transition
probability $p(\ph_f,t_f|\ph_i,t_i)$ in the weak-noise limit. 

We have hence to solve a
variational problem for which the initial and final point of the 
trajectory are not fixed. Instead they are required to lie on two
known functions $\ps$ and $\pus$, respectively. If we perform the
optimization with respect to these locations we find
\begin{equation}
  \pa_t \phin(t_i)=F(\phin(t_i),t_i)\quad\text{and}\quad
  \pa_t \phin(t_f)=F(\phin(t_f),t_f).
\end{equation}
Hence the optimal transition trajectory $\phin(t)$ must be {\em parallel} to
the stable and unstable orbit at the contact points. This in turn
implies $t_i=-\infty$ and $t_f=\infty$. The optimal transition
trajectory $\phin(t)$ hence starts out at $t_i=-\infty$ at the stable orbit,
moves for a very long time near to it, changes then in a rather short
time interval to the immediate neighbourhood of the unstable orbit 
(hence the name ``instanton'')
which it finally reaches at $t_f=\infty$ \cite{LRH}. 

{F}rom the minimization of the action $S[\ph (\cdot)]$ with respect to
the functions $\ph (t)$ we get the usual
Euler-Lagrange equation for the minimizing instanton $\phin (t)$
which for the action given by (\ref{action})
takes the form
\begin{equation}\label{ELE1d}
    \pa_t^2 \phin (t)=\pa_t F(\phin (t),t) + 
             F(\phin (t),t)\,\pa_\ph F(\phin (t),t) \ .
\end{equation}

Choosing for the time dependence of $H_y(t)$ the example specified by
(\ref{foft1}) we have solved (\ref{ELE1d}) numerically as 
a boundary value problem for a system of ordinary differential
equations. The initial and final points were chosen slightly off the
stable and unstable orbit, respectively. If the difference between the
initial and final time, $t_f-t_i$, is larger than several periods
of the external driving, the solution hardly depends on the
precise value of these deviations from the deterministic
orbits. Having obtained $\phin(t)$, the corresponding value of the
action $S$ can be calculated from (\ref{action}). 

\begin{figure} 
  \includegraphics[width=0.4\columnwidth]{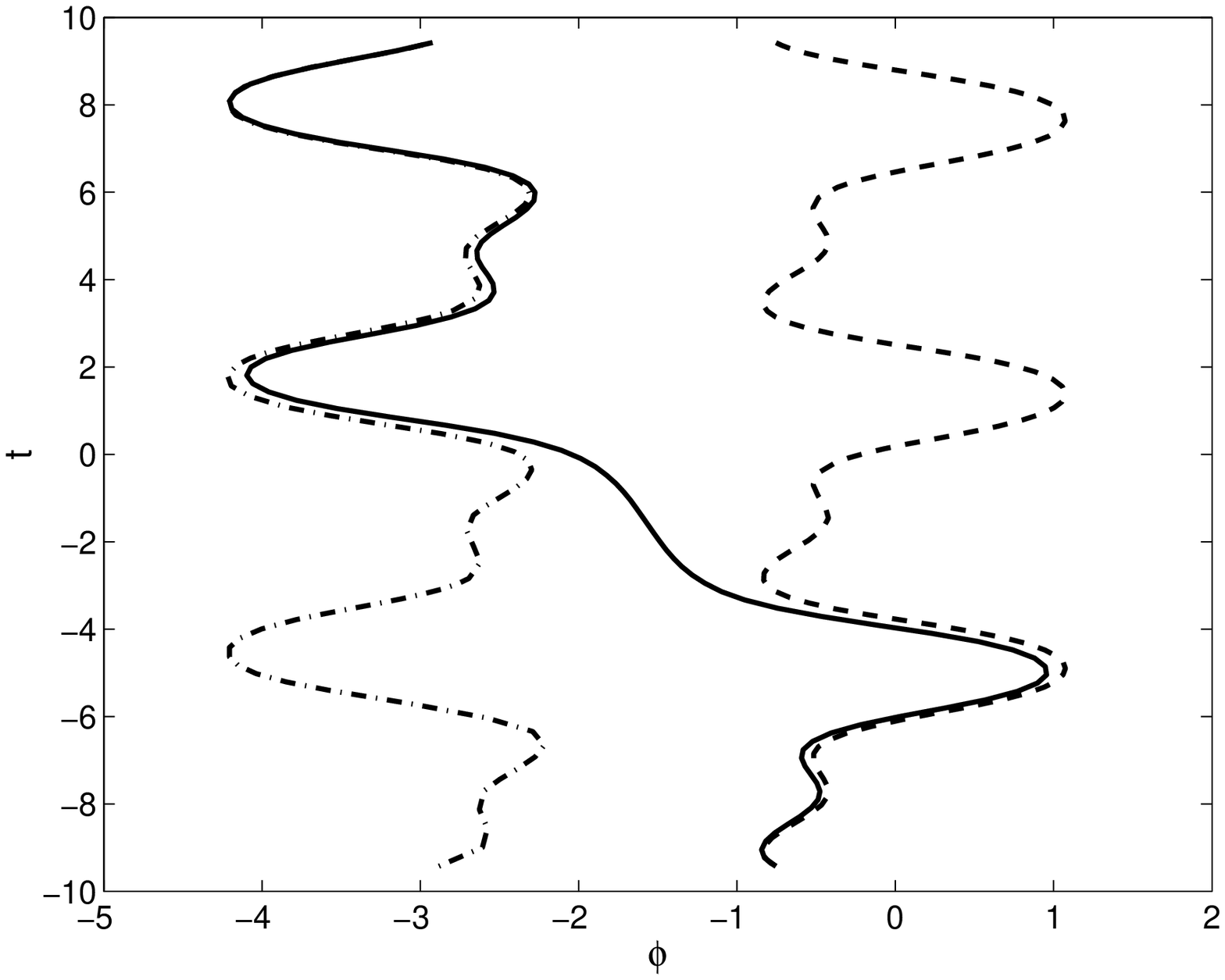}  
\hskip 0.5cm
  \includegraphics[width=0.4\columnwidth]{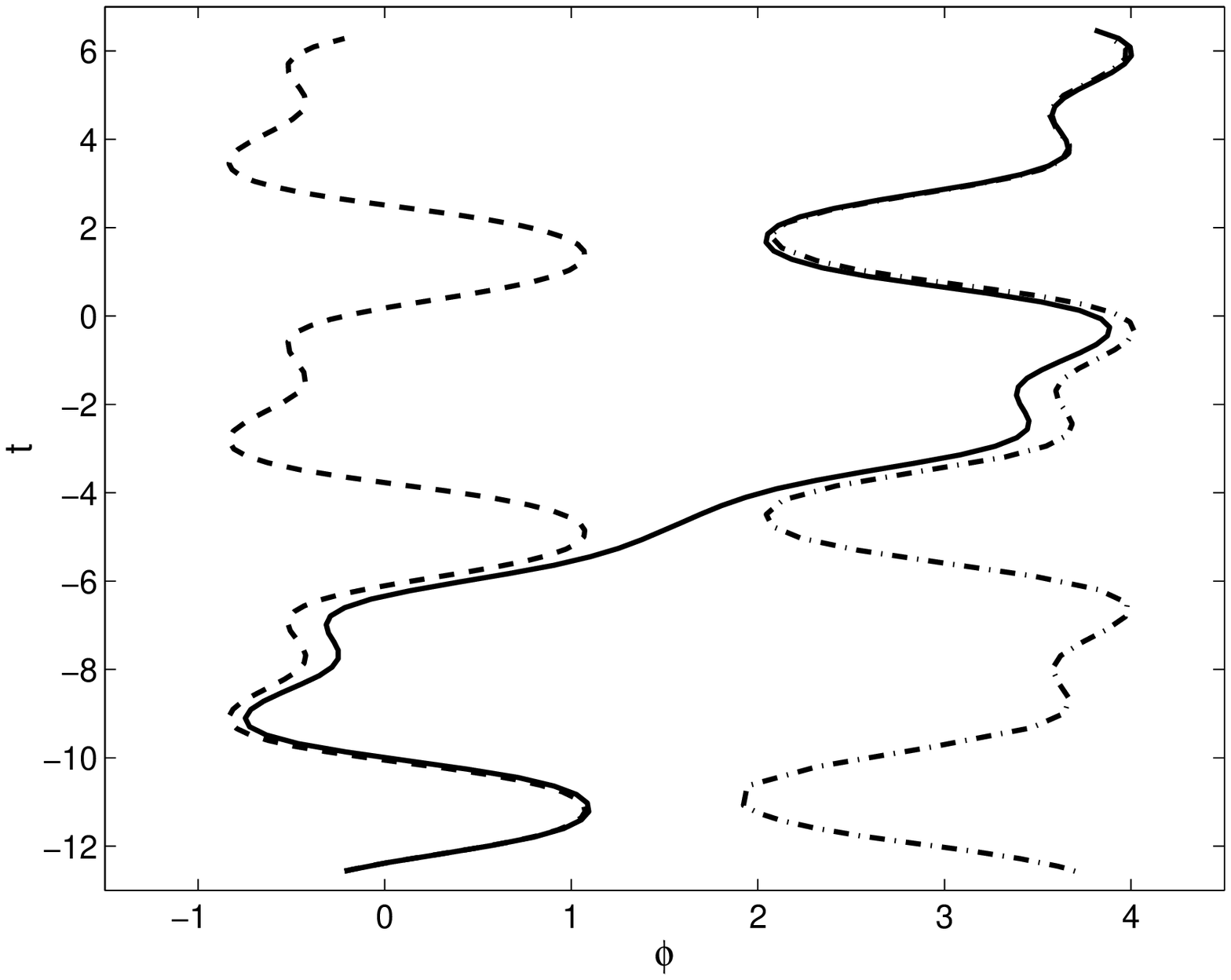}  
  \caption{\label{fig:inst1d} Numericaly determined
    instanton trajectories (full lines)
    describing the dominating weak noise stochastic transitions between stable
    (dashed lines) and unstable (dashed-dotted lines) deterministic
    periodic orbits of the one-dimensional rotational dynamics of a
    ferrofluid particle described by (\ref{lan3}), (\ref{defF}), (\ref{foft1}). 
    The parameter
    values are $H_x=0.3,\,H_y^{(1)}=H_y^{(2)}=1$ and $\delta=0$.
    For the forward instanton, increasing $\ph$ by
    $2\pi$ (right), the value of the action is $S=0.21472$, for the
    backward instanton, decreasing $\ph$ by $2\pi$ (left), it is
    $S=0.30969$. Hence, a noise induced average 
    net rotation in the ``forward direction'' results in the weak noise limit,
    implying $\Nav >0$ according to (\ref{eq24}).}
\end{figure}

In fig.~\ref{fig:inst1d} the resulting instanton trajectories $\phin(t)$
corresponding to backward and forward transitions are displayed. They 
show the qualitative behaviour discussed above. Note, however, that only
the noise induced transition from the stable to the unstable orbit is
shown. The complete transitions giving rise to $\ph\mapsto\ph\pm 2\pi$
comprises also the subsequent relaxations from the unstable orbit to
the next stable one. Consequently the stochastic transitions show
substantially more structure than the rather poor caricatures used to
represent them in Fig.~\ref{fig:fig1}. 

The corresponding values of the action are $S=0.21472$ for the forward
instanton increasing $\ph$ by $2\pi$ and $S=0.30969$ for the backward
instanton decreasing $\ph$ by $2\pi$. Hence, at least in the weak
noise limit, forward transitions are more likely (their action is
smaller) than backward transitions and correspondingly for the
parameter set specified in the caption of Fig.~\ref{fig:inst1d} we
find a noise driven rotation in the direction of increasing $\ph$. 
The mismatch between forward and backward transition clearly
demonstrates the operation of the ratchet effect in our system. 
With the relevant transition probabilities being of order
$\exp(-O(1)/D)$ the effect is however very weak and in particular
rapidly disappears with $D\to 0$.

\subsection{Numerical solution of the FPE}

For the numerical solution of the FPE describing the complete
two-dimensional dynamics for general values of the parameters it is
convenient to expand the probability distribution $P(\te,\ph,t)$ in
spherical harmonics $Y_l^m(\te,\ph)$ \cite{AbSt}  
\begin{equation}\label{Pexp}
  P(\te,\ph,t)=\sum_{l=0}^\infty \sum_{m=-l}^l
          A_{l,m}(t)\,Y_l^m(\te,\ph)\ , 
\end{equation}
with so far unknown time dependent complex expansion coefficients
$A_{l,m}(t)$. Since $P(\te,\ph,t)$ is real we have
\begin{equation}\label{Asym}
  A_{l,-m}(t)=(-1)^m A_{l,m}^*(t)\ ,
\end{equation}
with the star denoting complex conjugation. Moreover, from the
normalization of $P(\te,\ph,t)$ we find  
\begin{equation}\label{resA00}
  A_{0,0}(t)=\frac 1 {\sqrt{4\pi}}\ .
\end{equation}
Plugging the ansatz (\ref{Pexp}) into the FPE (\ref{FPE}), exploiting
(\ref{defpot}), and using well-known properties of the spherical
harmonics we find for $l\geq 1$ the following set of ordinary
differential equations 
\begin{align}\label{odeset}\nonumber
  \pa_t A_{l,m}(t)=
        &-\frac{H_x-i H_y(t)}{2}\Big(h_1(l-1,m-1)\,A_{l-1,m-1}(t)
                         +h_2(l+1,m+1)\,A_{l+1,m-1}(t)\Big)\\\nonumber
        &+\frac{H_x+i H_y(t)}{2}\Big(h_1(l-1,-m-1)\,A_{l-1,m+1}(t)
                         +h_2(l+1,-m-1)\,A_{l+1,m+1}(t)\Big)\\
        &-D\,l(l+1)\, A_{l,m}(t)
\end{align}
where we have introduced the auxiliary factors 
\begin{align}
  h_1(l,m)&=(l+2)\,\sqrt{\frac{(l+m+1)(l+m+2)}{(2l+1)(2l+3)}}\\
  h_2(l,m)&=(l-1)\,\sqrt{\frac{(l-m)(l-m-1)}{(2l+1)(2l-1)}}.
\end{align}
Moreover, we find for the ensemble averages of the components of $\e (t)$
from (\ref{defe})
\begin{eqnarray}
    \la e_x (t) \ra = \la \sin\theta(t)\, \cos\ph(t) \ra 
    = -\sqrt{\frac{8\pi}{3}}\,\Re\, A_{1,1}(t)
    \\
    \la e_y (t) \ra = \la \sin\theta(t)\, \sin\ph(t) \ra 
    =\sqrt{\frac{8\pi}{3}}\,\Im\, A_{1,1}(t)
    \\
    \la e_z (t) \ra = \la \cos\theta(t) \ra 
    =\sqrt{\frac{4\pi}{3}}\,    A_{1,0}(t)
\end{eqnarray}
where $\Re z$ and $\Im z$ denote the real and imaginary part of the
complex number $z$, respectively. 
For the $z$-component of the ensemble averaged but still time-dependent
magnetic torque $\la N_z\ra$ this implies via (\ref{defmagtor})
\begin{equation}\label{torexp}
  \la N_z\ra=-\sqrt{\frac{8\pi}{3}}\,
        \big(H_y(t)\,\Re\, A_{1,1}(t)+H_x \,\Im\,A_{1,1}(t)\big) \ .
\end{equation}

We have used two somewhat different procedures to explicitly calculate
this torque. The first is more general, the second is slightly more
efficient for the special case of the time dependence (\ref{foft1}). 

In the first method we solve the system (\ref{odeset}) of ordinary
differential equations for the expansion coefficients $A_{l,m}(t)$ 
with $l\leq l_{\mathrm{max}}$ by a standard routine. We have used
values of $l_{\mathrm{max}}$ up to 25 but found typically 
$5\leq l_{\mathrm{max}}\leq 10$ to be sufficient. Starting with some
initial condition, we integrate the equation until the periodic time
dependence of the solution is reached, which was usually the case after
at most two periods of the external field. We then integrate further
for one period of the external driving, use the result for
$A_{1,1}(t)$ in (\ref{torexp}), and calculate the remaining time
average numerically. Some care is needed 
here since this time-averaged torque $\Nav$ is typically 2-3 orders of
magnitude smaller than the typical time-dependent values of $N(t)$. In the
time average we are hence subtracting numbers of equal size from each
other giving rise to well-known problems of numerical accuracy. The
advantage of this method is that it works for any numerically sensible
time dependence of the magnetic field $H_y(t)$ in (\ref{deffield}).

The second method builds on
the fact that the long-time solution of the FPE will be periodic in
time with the period $2\pi$ of the external driving. It is hence
useful to expand the time-dependent expansion coefficients
$A_{l,m}(t)$ in (\ref{Pexp}) into Fourier modes with respect to time
\begin{equation}\label{Fourier}
  A_{l,m}(t)=\sum_{s=-\infty}^\infty \tilde{A}_{l,m,s}\,e^{ist}.
\end{equation}
Instead of a system of ODE's for the coefficients $A_{l,m}(t)$ we find
now from the FPE (\ref{FPE}) a system of {\em algebraic} equations for
the  coefficients $\tilde{A}_{l,m,s}$. Moreover, the time averaged
torque $\Nav$ can be readily expressed as function of some of these
coefficients by means of (\ref{torexp}). 
The advantages of this method are that effectively we are
directly dealing with the stationary solution of the FPE, {\it i.e.} no
initial conditions are 
necessary and no equilibration process must be simulated, and
that the time average of the torque needs not to be performed
numerically. The disadvantage is that we have to specify $H_y(t)$ in
order to find the algebraic system for the $\tilde{A}_{l,m,s}$ and
that for general $H_y(t)$ this system will be rather dense with respect
to the index $s$. 

For the special time dependence (\ref{foft1}) the situation is somewhat
more gratifying since $H_y(t)$ involves 
only two Fourier components. Hence there are only couplings between
coefficients $\tilde{A}_{l,m,s}$ differing in their $s$-value by
at most 2 and the result for $\Nav$ is a linear combination of just 5
different coefficients $\tilde{A}_{l,m,s}$. This makes the numerical
analysis rather fast. Both methods must of course yield the same
results when applied to the same $H_y(t)$. We have frequently used both
to verify our numerical findings. 

Using the numerical methods described above we have verified all the
predictions derived in section \ref{sec:symm} on the basis of symmetry
arguments. In particular we find $\Nav=0$ if $H_x=0$ in accordance
with (\ref{4a}) and $\Nav=0$ for $H_y(t)=\cos(t)$, {\it i.e.} for 
(\ref{foft1}) with $H_y^{(2)}=0$, or 
$H_y(t)=H_y^{(1)}\cos(t)+ H_y^{(2)}\cos(3t)$ in accordance
with (\ref{4b}). 

\begin{figure} 
  \includegraphics[width=0.4\columnwidth]{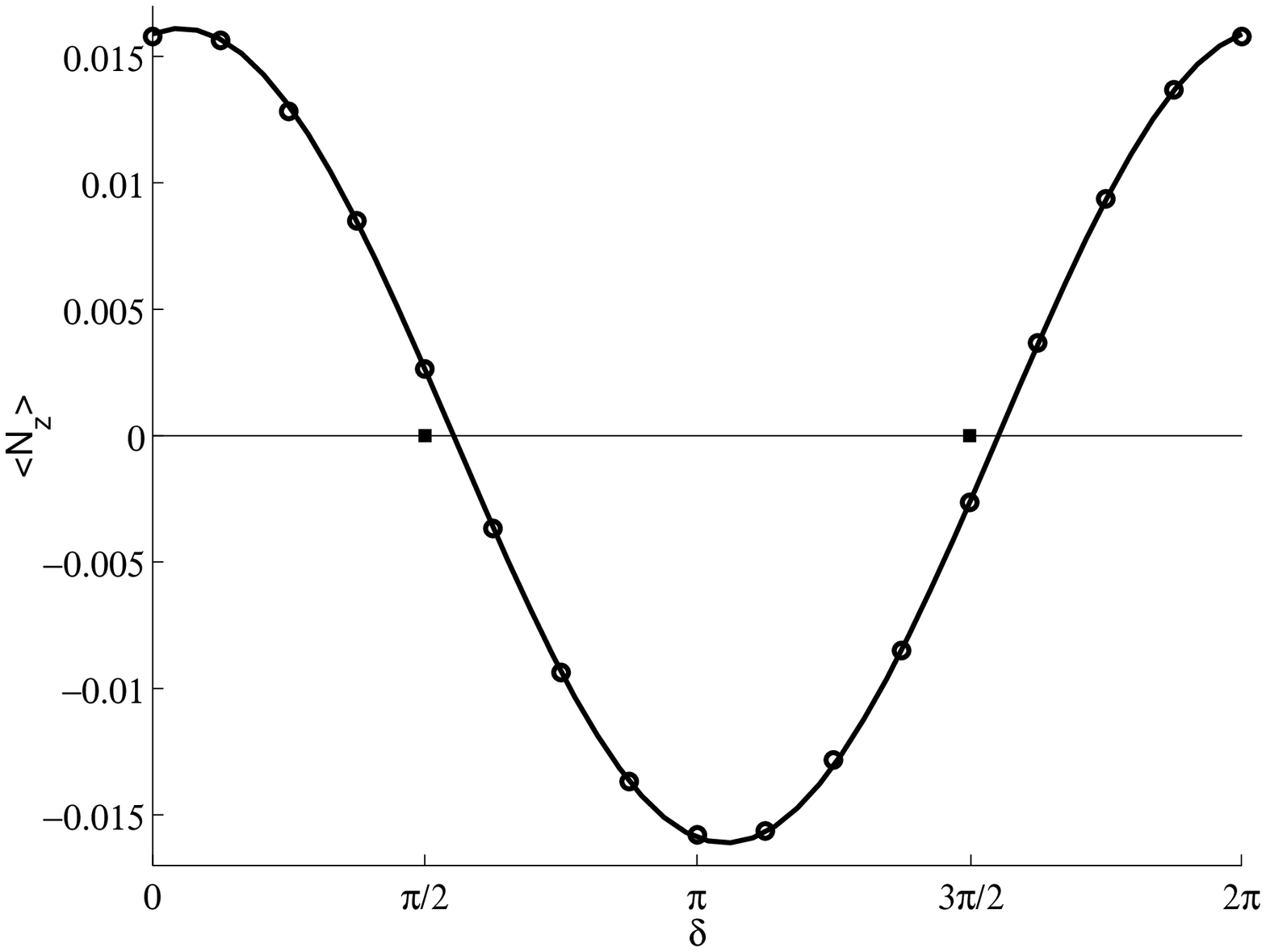}  
\hskip 0.5cm
  \includegraphics[width=0.4\columnwidth]{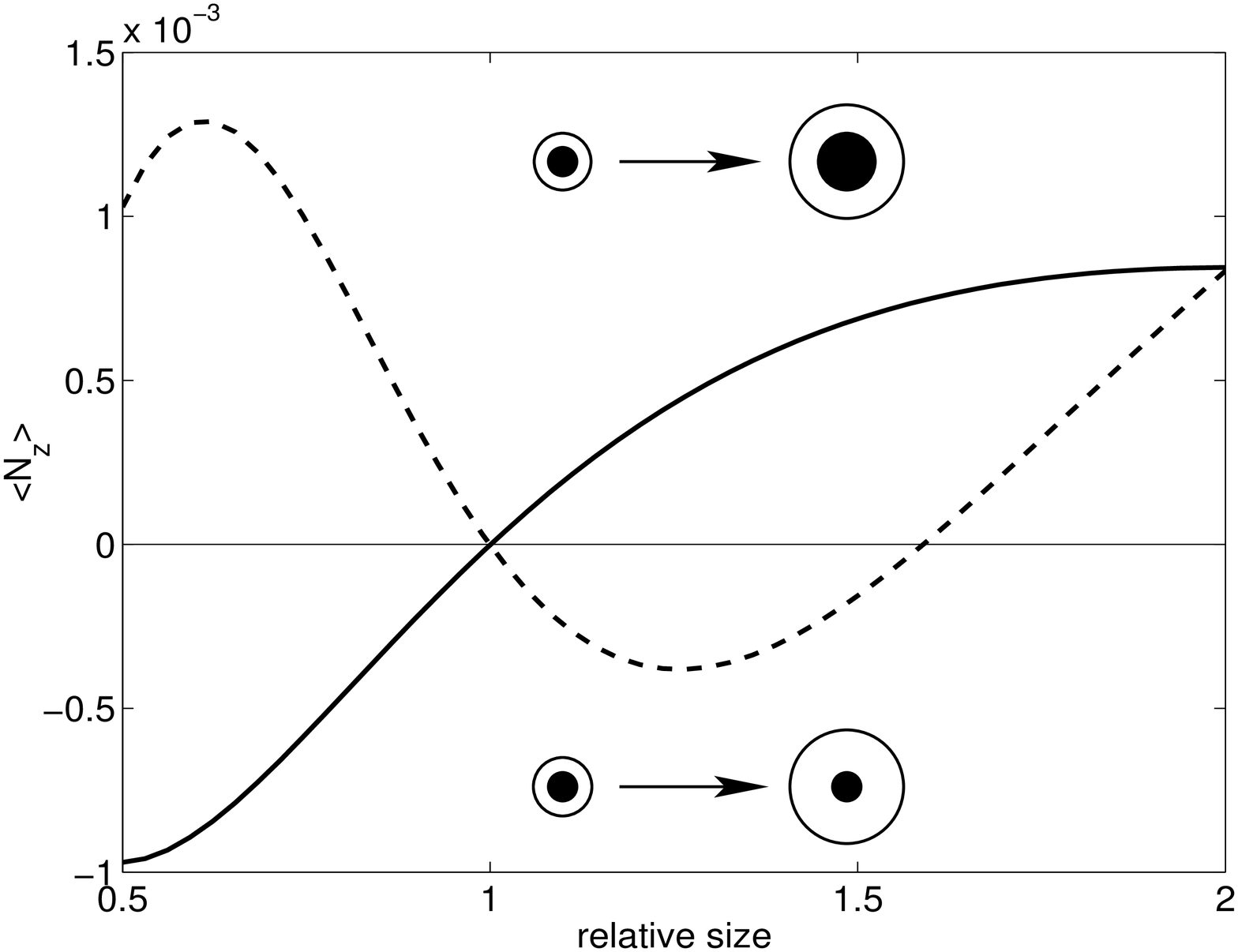}  
  \caption{\label{fig:numres} Left: Averaged torque $\Nav$ for the
    time modulation (\ref{foft1}) with 
    $H_x=0.3,\,H_y^{(1)}=H_y^{(2)}=1$ and $D=0.2$ as function of the
    phase angle $\delta$ as obtained from the numerical solution of
    the FPE (full line). Symbols are results from simulations of the
    Langevin equations (\ref{lan1}) and (\ref{lan2}) as reported in
    \cite{RaEn} with statistical errors being smaller than the symbol
    size. The squares on the horizontal axis are the points
    $(\pi/2,0)$ and $(3\pi/2,0)$ at which the current reversal would
    occur in the reduced model (\ref{lan3}). Right: Dependence of 
    $\Nav$ on the relative particle size for the same values of the 
    parameters and $\delta$ fixed to $\delta_0\simeq 1.73$, the first
    point of current inversion in the left panel. The full line
    corresponds to a proportional change of the magnetic and
    hydrodynamic radii (top sketch), the dashed line is for a change of the
    hydrodynamic radius only (bottom sketch).}
\end{figure}

In Fig.~\ref{fig:numres} we give some examples for current
inversions in our system. The left panel shows the averaged torque
$\Nav$ as function of the phase angle $\delta$
for a time dependent magnetic field $H_y(t)$  
of the form (\ref{foft1}). The results are  
consistent with (\ref{Nofdel}). Moreover, the values of
$\delta$ at which the torque changes sign are clearly different from 
$\pi/2$ and $3\pi/2$ indicated by the squares on the horizontal
axis. Hence the numerical solution verifies the statement of
section~\ref{sec:symm} that the current reversal in the full two
dimensional model (\ref{lan1}), (\ref{lan2}) occurs at values
of $\delta$ which are not fixed by any symmetry but rather depend
in a complicated manner on all the other parameters of the problem. 

This is further corroborated by the right panel of Fig.~\ref{fig:numres},
depicting the torque $\Nav$ as function of the relative particle size for
$\delta=\delta_0$. There are two different ways to scale the particle size
that make sense in a ferrofluid. In the first, both the ferromagnetic
core of the particle as well as the polymer coating are scaled with the same
factor. This gives rise to a proportional change of the magnetic
moment $\mu$ and the particle volume $V$. In the dimensionless units
adopted this implies unchanged fields $H_x$ and $H_y(t)$ and a
noise intensity $D$ rescaled according to (\ref{defD}). In the other
case, only the polymer coating is scaled and therefore
only the hydrodynamic radius of the particle changes. Hence the
magnetic moment remains the same whereas the volume $V$
changes. Correspondingly the dimensionless fields $H_x$ and $H_y(t)$
and the noise intensity $D$ change by the same factor. In view of
(\ref{FPE}) this can be absorbed in a rescaled time which in turn is
equivalent to a changed frequency $\om$ of the external field. 
As anticipated in section~\ref{sec:symm}, we find current inversions in
both cases.

\subsection{Effective field approximation}
\label{sec:efff} 

It is very helpful to have some analytical expression for the average
torque $\Nav$ since the dependencies on the various parameters of the
system can then be identified much more directly. In the present and
subsequent subsections we discuss two approximate methods to calculate
the time averaged torque in our system. 

The first one employs the so-called effective field method to
approximately solve the FPE. It has been a standard tool in
the theory of ferrofluids for many years \cite{MRS}. Two main
ingredients are necessary. First, let us recall that the stationary
(equilibrium) solution of the FPE (\ref{FPE}) for a 
{\em time independent} homogeneous magnetic field $\Ha$ is of the form
\begin{equation}\label{Pnot}
  P^{(0)}(\e)=\frac{H}{4\pi\,D\,\sinh (H/D)}\;
             \exp(\frac{\e\cdot\Ha}{D})\ ,
\end{equation}
where $H$ denotes the modulus of the magnetic field, $H=|\Ha|$.
Averages with this distribution are easily calculated, in particular
we find the well-known result
\begin{equation}\label{enot}
  \la\e\ra^{(0)}=
        L(H/D)\, \Ha/H
\end{equation}
with the Langevin function 
\begin{equation}\label{defL}
  L(x)=\coth(x)-1/x \ .
\end{equation}
Second, for general, time-dependent fields 
$\Ha(t)$ the FPE (\ref{FPE}) can be used to derive
the {\em exact} equation  
\begin{equation}\label{exequ}
  \pa_t \la \e \ra+2D \la \e \ra =
       -\la \e \times (\e \times \Ha)\ra
\end{equation}
for the time evolution of the average $\la\e\ra$. Note that this
equation for the first moment of $P(\e,t)$ has the usual flaw of being
not closed but involving higher moments of the probability distribution.

The central idea of the effective field approximation is now to assume
that the solution $P(\e,t;\Ha(t))$ of the FPE for a general time 
dependent magnetic field $\Ha(t)$ can be written as equilibrium
distribution for some a-priori unknown {\em effective field}
$\Ha_e(t)$, {\it i.e.} $P(\e,t;\Ha(t))=P^{(0)}(\e;\Ha_e(t))$. The average on
the r.h.s. of (\ref{exequ}) can then be performed and this equation
gives rise to an evolution equation for the effective field $\Ha_e$ of
the form \cite{MRS} 
\begin{equation}\label{efff}
\pa_t\big(\frac{L(H_e/D)}{H_e}\,\Ha_e\big)=
   -2D\frac{L(H_e/D)}{H_e}\,(\Ha_e-\Ha)
   -\frac{H_e-3 D L(H_e/D)}{H_e^3}\,(\Ha_e\times(\Ha_e\times\Ha))
\end{equation}
Given $\Ha(t)$, this is a closed equation for the time evolution of
$\Ha_e(t)$. Having obtained $\Ha_e(t)$ we get the desired approximate
result for $\la\e\ra$ from (\ref{enot})
\begin{equation}\label{eefff}
  \la\e\ra_e=L(H_e/D)\, \Ha_e/H_e \ .
\end{equation}
The solution of the partial differential equation (\ref{FPE}) is hence
replaced by the solution of the set of three coupled ordinary differential
equations (\ref{efff}). There are other, equivalent forms of
(\ref{efff}), see {\it e.g.} \cite{MuLi}. Exploiting (\ref{efff}), it
is also possible to derive the following equivalent closed equation
for $\la\e\ra$ \cite{EMRJ} 
\begin{equation}\label{efff1}
  \pa_t\la\e\ra_e=-(1-\frac{D\la e\ra_e}{H_e})(\Ha_e-\Ha)
     +(1-3D\frac{\la e\ra_e}{H_e})
      \big(\frac{\la\e\ra_e}{\la e\ra_e}\cdot(\Ha_e-\Ha)\big)\,
      \frac{\la\e\ra_e}{\la e\ra_e} \ ,
\end{equation}
where $\la e \ra_e$ denotes the modulus of $\la\e\ra_e$ and the
effective field $\Ha_e$ is to be expressed as function of $\la\e\ra_e$
by the inverse of (\ref{eefff}). Note the difference between 
$\la e \ra_e=|\la\e\ra_e|$ and $\la |\e|\ra_e\equiv 1$. 

\begin{figure} 
  \includegraphics[width=0.4\columnwidth]{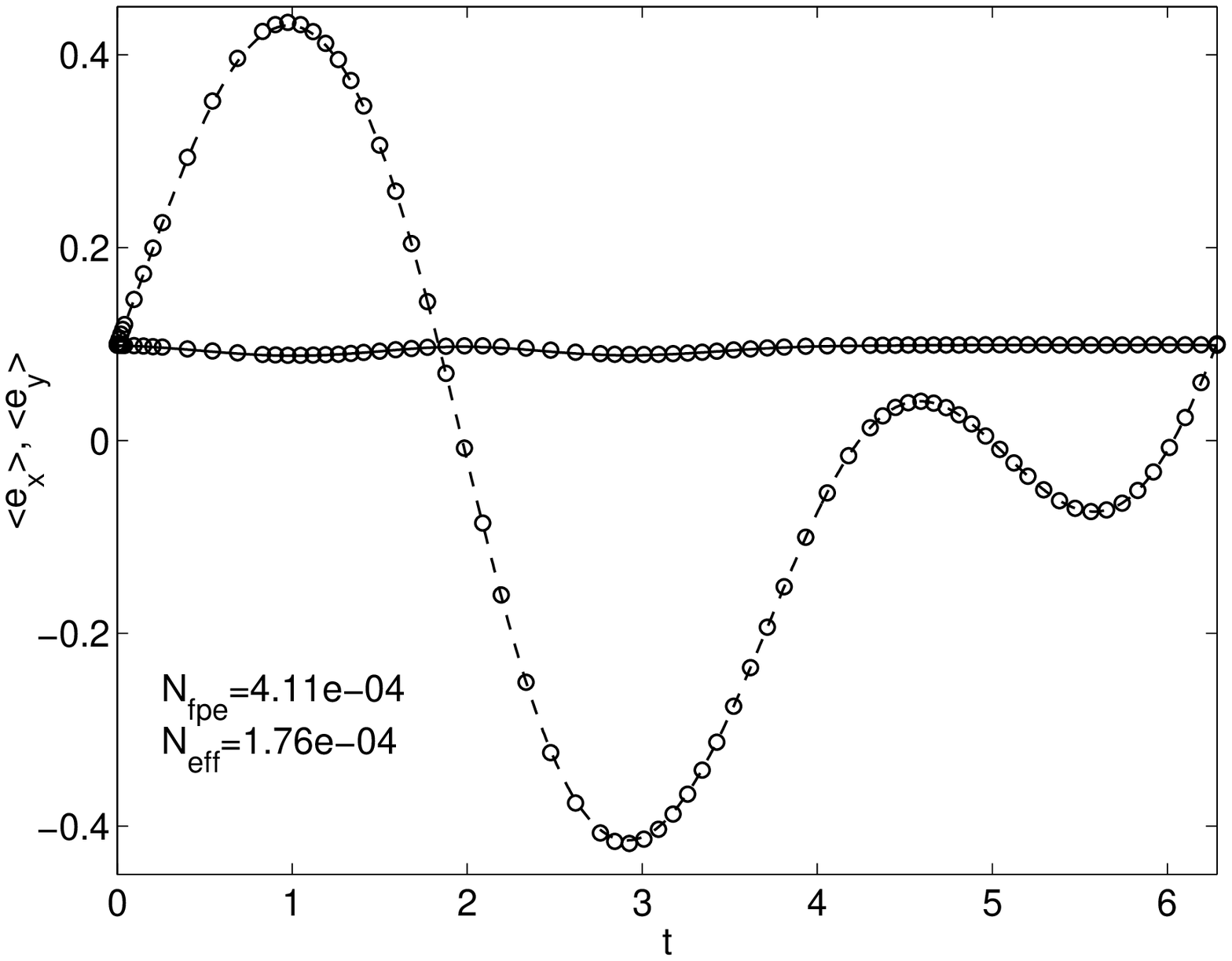}  
\hskip 0.5cm
  \includegraphics[width=0.4\columnwidth]{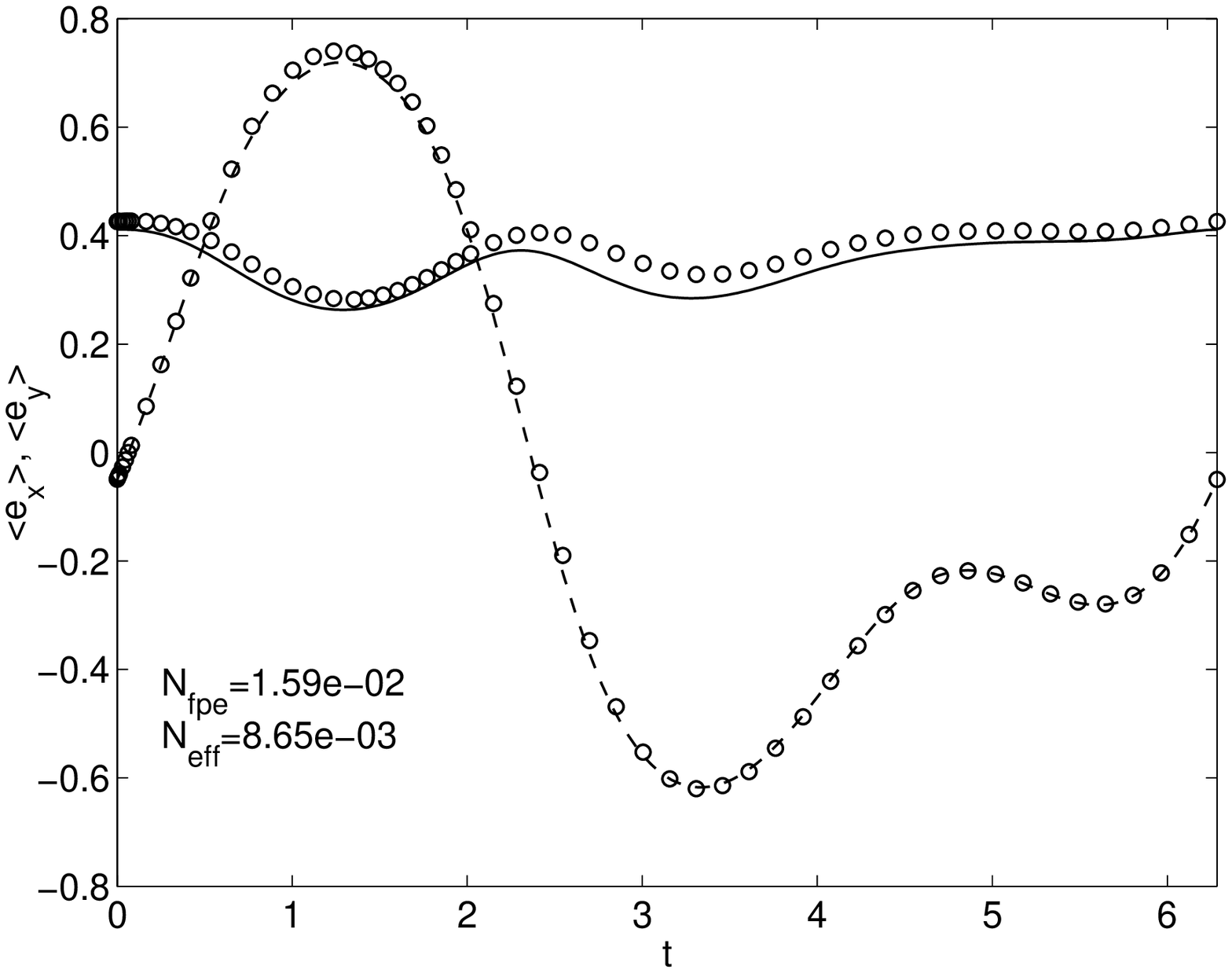}  
  \caption{\label{fig:efff} Horizontal components of the average
    orientation $\la\e\ra$ of the particle as function of time for one
    period of the external magnetic field. The lines show the results
    for the $x$-and $y$-component of $\la\e\ra$ (full and dashed line,
    respectively) from the numerical solution of the Fokker-Planck
    equation (\ref{FPE}) with a time dependence of the external field as
    given in (\ref{foft1}). The symbols are the corresponding results
    from the effective field theory. The parameter values are 
    $H_x=0.3,\,H_y^{(1)}=H_y^{(2)}=1,\,\delta=0$ as well as $D=1$
    (left) and $D=0.2$ (right). The corresponding values for the time
    averaged torque $\Nav$ are also displayed.}
\end{figure}

In general, (\ref{efff}) can only be solved numerically. In
Fig.~\ref{fig:efff} we give a comparison between results obtained in
this way for the time dependence (\ref{foft1}) and the corresponding
outcome of the numerical solution of the FPE. It is clearly seen that
the effective 
field approximation yields rather accurate results for the time
dependent orientation of the particle and, equivalently, for the time
dependent magnetization of the ferrofluid. On the other hand the
results for the time averaged torque $\Nav$ differ from the
numerically exact values by a factor between 2 and 3. The reason for
this discrepancy lies in the fact that $\Nav$ is typically 2-3 orders
of magnitude smaller than typical values of the time dependent torque 
$\la N_z\ra$. In calculating the time average of the torque we hence
subtract quantities of comparable order and so amplify the inaccuracies of the
effective field approximation from a few percent to a few hundred percent.

In order to obtain an analytical expression for $\Nav$ within the
framework of the effective field approximation, it is useful to
consider the case of small values of $H/D$ in which the Langevin
function $L(x)$ or equivalently its inverse may be approximated by the
first terms of the respective power series expansion. Using the lowest order
approximation $L(x)\simeq x/3$ we have $H_e\simeq 3D \la e \ra_e$ and
from (\ref{efff1}) it then follows that
\begin{equation}
  \pa_t \la\e\ra_e + 2 D \la\e\ra_e =2\Ha/3 \ .
\end{equation}
For large $t$ we therefore find
\begin{align}\label{efffpert1}
  \la e_x\ra_e &\to \frac{2 H_x}{3D}\\\label{efffpert2}
  \la e_y\ra_e &\to \frac{2 H_y^{(1)}}{3}\frac{2D\cos t + \sin t}{4D^2+1}
      +\frac{H_y^{(2)}}{3}
        \frac{D\sin(2t+\delta)-\cos(2t+\delta)}{D^2+1} \ .
\end{align}
Hence $\la e_x\ra_e$ tends asymptotically to a constant whereas 
$\la e_y\ra_e$ contains only oscillating parts. From (\ref{defmagtor}) we
may thus infer that the torque $\la N_z \ra$ harmonically oscillates and
therefore its time average vanishes, $\Nav=0$. 

To get a non-trivial result for $\Nav$ we have to push the
expansion of the Langevin function further. In other words, the
non-linearity of the magnetization curve is essential to get a
non-zero average torque. If the next 
term in the expansion is included, $L(x)\simeq x/3-x^3/45$, we already
obtain an approximation for $L(x)$ that is accurate to within one per
cent for the experimentally relevant values of $x$ which are generally
less than $1$. Calculating the corrections to (\ref{efffpert1}) and 
(\ref{efffpert2}) induced by these higher order terms we finally find
after some algebra the following result for the average torque
\begin{equation}\label{effresN}
 \Nav_e=\frac{H_x\,(H_y^{(1)})^2\,H_y^{(2)}}{30}\;
   \frac{\cos\delta+2 D \sin\delta}
        {(1+D^2)(1+4 D^2)^2} \ .
\end{equation}
This result is in accordance with our symmetry considerations of
section \ref{sec:symm} since it shows that both a non-zero field
component in $x$-direction and an even higher harmonic in the
time dependence of $H_y(t)$ are essential for a non-zero torque to occur,
cf. (\ref{4a}), (\ref{4b}). 


\subsection{Perturbative solution of the FPE}\label{sec:pert}

In the last subsection we saw that the effective field approximation
alone is not sufficient to get an explicit analytical expression for
the time averaged torque $\Nav$ in our system. Additionally we had to
expand the Langevin function in its argument $H/D$ which amounts to a
small field or equivalently large noise expansion. Moreover, the
approximate result for the torque was typically at variance with the
numerically exact result by a factor of about 2. 

It is therefore tempting to avoid the effective field approximation
altogether and to use a perturbation expansion in the ratio of
magnetic field strength and noise intensity right in the Fokker-Planck
equation (\ref{FPE}). This is conveniently accomplished by making the
formal substitution $\Ha\mapsto\eps\Ha$ in (\ref{FPE}) and by expanding
the solution of the FPE in a power series in $\eps$  
\begin{equation}\label{Ppert}
  P(\te,\ph,t)=P^{(0)}(\te,\ph,t)+\eps\,P^{(1)}(\te,\ph,t)
               +\eps^2\,P^{(2)}(\te,\ph,t)+\eps^3\,P^{(3)}(\te,\ph,t)
               +\dots\quad.
\end{equation}
At the end of the calculation $\eps$ is put equal to one. 
It will turn out that we have to use the expansion up to third order
in $\eps$ to get a non-zero result for the average torque $\Nav$.
Since the potential $U(\te,\ph,t)$ is of order $\eps$ we find in 
zeroth order 
\begin{equation}\label{resP0}
  \lim_{t\to\infty}P^{(0)}(\te,\ph,t)=\frac 1 {4\pi}\quad,
\end{equation}
describing the stationary distribution of pure rotational
diffusion. Using the expansion (\ref{Ppert}) in (\ref{FPE}) we get a
system of linear inhomogeneous partial differential equations for the
various $P^{(n)}(\te,\ph,t)$ that can be solved in principle one
after the other. 

This procedure is, however, rather cumbersome. It can be somewhat
simplified by using the exact equation (\ref{exequ}), the r.h.s of
which is of order $\eps$. Hence to find $\la\e\ra$ from this equation
to order $\eps^n$ it is sufficient to calculate the average on
the r.h.s. with an expression for $P(\te,\ph,t)$ accurate to order
$\eps^{(n-1)}$. In our case the lowest order non-zero result for $\Nav$ is
$\Ord (\eps^4)$, hence we need $\la\e\ra$ to order $\eps^3$
(cf. (\ref{defmagtor}), (\ref{defe})) and therefore the second order
result for $P(\te,\ph,t)$ would be sufficient.  

It is, however, more convenient to use again the expansion
(\ref{Pexp}) of $P(\te,\ph,t)$ in terms of spherical harmonics. The
perturbation ansatz (\ref{Ppert}) then translates into an expansion of
the coefficients $A_{l,m}(t)$ in powers of $\eps$
\begin{equation}
  A_{l,m}(t)=A^{(0)}_{l,m}(t)+\eps\, A^{(1)}_{l,m}(t)
      +\eps^2\, A^{(2)}_{l,m}(t) + \eps^3\, A^{(3)}_{l,m}(t)+\dots  
\end{equation}
The ODE's (\ref{odeset}) couple only coefficients $A_{l,m}(t)$ which
differ in both $l$ and $m$ by 1. Moreover, we know from (\ref{resA00})
$A_{0,0}(t)=1/\sqrt{4\pi}=\Ord (1)$ and hence  $A_{l,m}(t)=\Ord
(\eps^l)$. As shown by (\ref{torexp}) the torque is determined by
$A_{1,1}(t)$ alone. We hence need an expression for $A_{1,1}(t)$ 
correct up to order $\eps^3$.  

\begin{figure}[htb]
  \includegraphics[height=6.4cm]{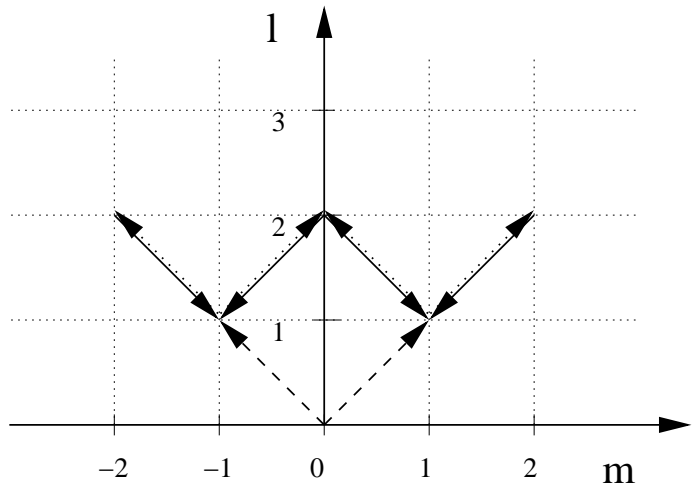}  
  \caption{\label{fig:pertdiag} Diagrammatic sketch of the
    perturbative determination of the expansion
    coefficients $A_{l,m}$ in (\ref{Pexp}). As shown by the arrows
    only $A_{l,m}$ differing in both $l$ and $m$ by 1 are coupled by
    equations (\ref{odeset}). In a first step we start
    from $A_{0,0}$ as given in (\ref{resA00}) and obtain the $\Ord (\eps)$
    terms of $A_{1,\pm 1}$ (dashed arrows). Next  $A_{2,0}$ and
    $A_{2,\pm 2}$ are determined to order $\eps^2$ (dotted
    arrows). Finally these are used to get the $\Ord (\eps^3)$
    contributions to $A_{1,1}$ (full arrows) from which the torque can
    be extracted to order $\eps^4$ using (\ref{torexp}). All other
    $A_{l,m}$ are irrelevant for $\Nav$ at this order in $\eps$. Note
    also that the left half of the diagram is redundant due to the
    symmetry property (\ref{Asym}).}
\end{figure}

To this end we determine from (\ref{odeset}) and (\ref{resA00}) first 
the $\Ord (\eps)$ terms of $A_{1,m}(t)$. With these we calculate the
necessary $A_{2,m}(t)$ to order $\eps^2$ which in turn are used to 
determine the $\Ord (\eps^3)$ terms in $A_{1,1}(t)$. The procedure is
sketched in fig.~\ref{fig:pertdiag}. Note that it is sufficient to
calculate coefficients $A_{l,m}(t)$ with $m \geq 0$ because of the
symmetry property (\ref{Asym}). 

In this way we first get from the $\Ord(\eps)$ equations
\begin{align}\nonumber
  (\pa_t +2D) A^{(1)}_{1,1}(t)
           &=-\frac 1 {\sqrt{6\pi}}\,\big(H_x-iH_y(t)\big)\\
  (\pa_t +2D) A^{(1)}_{1,0}(t)&=0.\nonumber
\end{align}
Since we are interested in the asymptotic solution valid for large $t$
we can put $A^{(1)}_{1,0}(t)\equiv 0$. Next we use the $\Ord(\eps^2)$
equations 
\begin{align}\nonumber
  (\pa_t +6D) A^{(2)}_{2,2}(t)&=
   -\frac 3 {\sqrt{5}}\,\big(H_x-iH_y(t)\big)A^{(1)}_{1,1}(t) \\\nonumber
  (\pa_t +6D) A^{(2)}_{2,0}(t)&= \sqrt{\frac 6 5}\,
       \Big(H_x\,\Re A^{(1)}_{1,1}(t)-H_y(t)\,\Im A^{(1)}_{1,1}(t)\Big)
\end{align}
to determine the relevant $A_{2,m}(t)$ to the desired order. Finally we
obtain $A^{(3)}_{1,1}(t)$ from 
\begin{equation}
  (\pa_t +2D) A^{(3)}_{1,1}(t)=
          -\frac 1 {\sqrt{30}} \big(H_x-iH_y(t)\big) A^{(2)}_{2,0} (t)
          +\frac 1 {\sqrt{5}} \big(H_x+iH_y(t)\big) A^{(2)}_{2,2}(t) \ . 
\end{equation}
The explicit expressions for the various coefficients $A^{(n)}_{l,m}(t)$
are rather long, the above system of equations is therefore
conveniently solved with the help of a computer algebra. 
Focusing on the specific time dependent field $H_y(t)$ from (\ref{foft1}),
exploiting (\ref{torexp}) and performing the time average
over one period of the external driving we finally get the following
lowest order perturbative result for the average torque 
\begin{equation}\label{pertresN}
 \Nav_p=\frac{H_x\,(H_y^{(1)})^2\,H_y^{(2)}}{40}\;
 \frac{9(1+29 D^2)\cos\delta+2 D (1+99 D^2)\sin\delta}
      {(1+D^2)(1+4D^2)(1+9D^2)(1+36D^2)} \ .
\end{equation}
The dependencies on the parameters $H_x,\,H_y^{(1)}$, and $H_y^{(2)}$
are identical to those of the effective field result (\ref{effresN}), whereas
the dependence on the noise strength $D$ is more complicated. 
One readily verifies that the symmetry properties (\ref{3a})-(\ref{3d})
are satisfied by (\ref{pertresN}). In particular, these
symmetries imply that the torque $\Nav$ must to be an odd function
of both the static field and the oscillating field. Moreover, 
symmetry reasons also imply \cite{Reirev} that in
linear order of the oscillating field one cannot expect a finite
$\Nav$. Rather, a coupling of several modes of the oscillating
field (harmonic mixing) is required, see also \cite{SeMa,GoHa,fla00}.
All together, these arguments explain that the first nontrivial
contribution must be (at least) of fourth order in the magnetic field
strength and that within a linear response theory one will
always find $\Nav = 0$ \cite{ala00}. On the other hand this does of
course not imply that the non-linear magnetization curve alone would 
``explain'' the effect. To understand the microscopic origin of the
angular momentum transfer from the external field to the particle one
really has to go to the description in terms of the FPE as
discussed above, see also \cite{Shcom,rep}.

\begin{figure}
  \includegraphics[width=0.4\columnwidth]{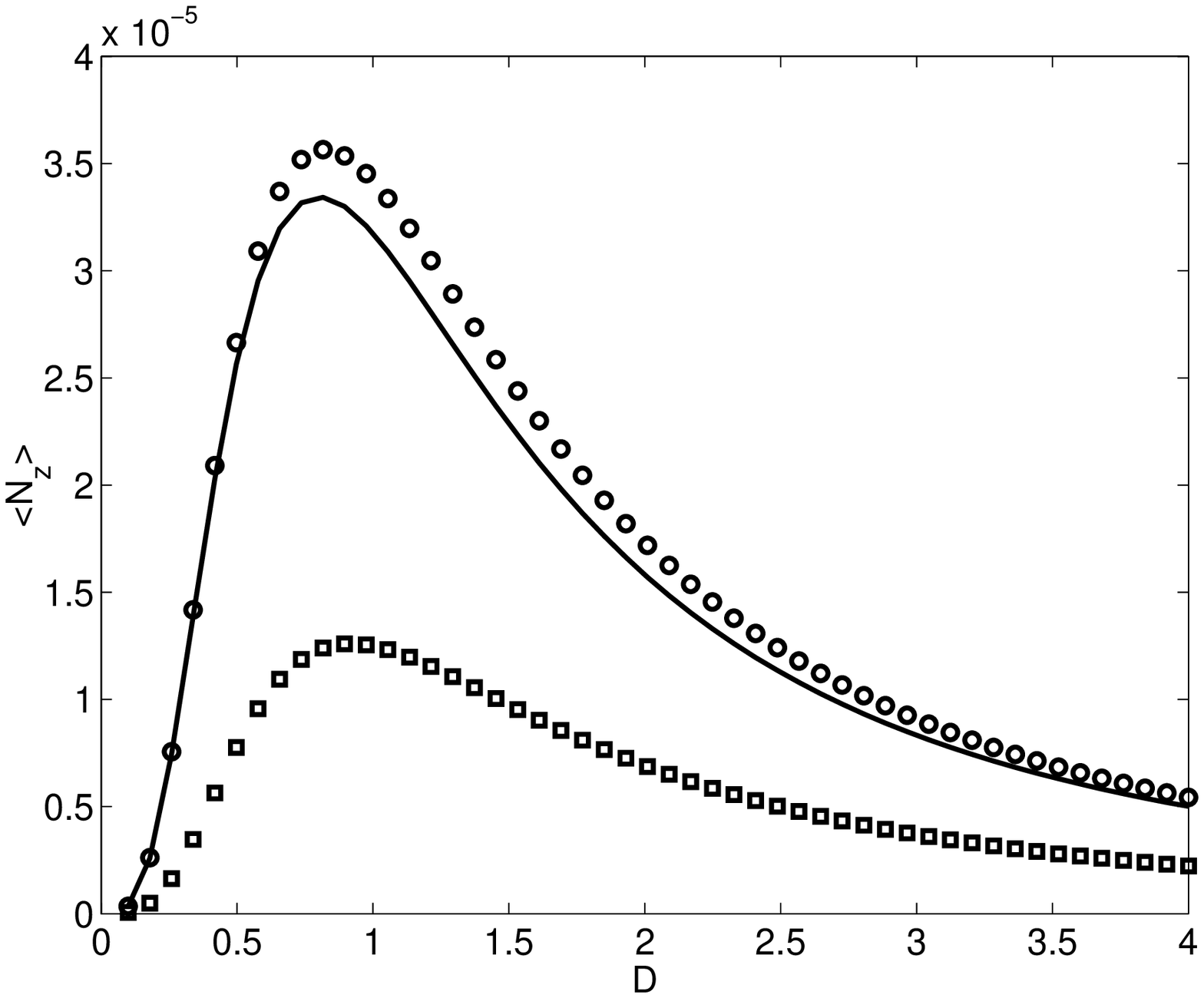}  
\hskip 2cm
  \includegraphics[width=0.4\columnwidth]{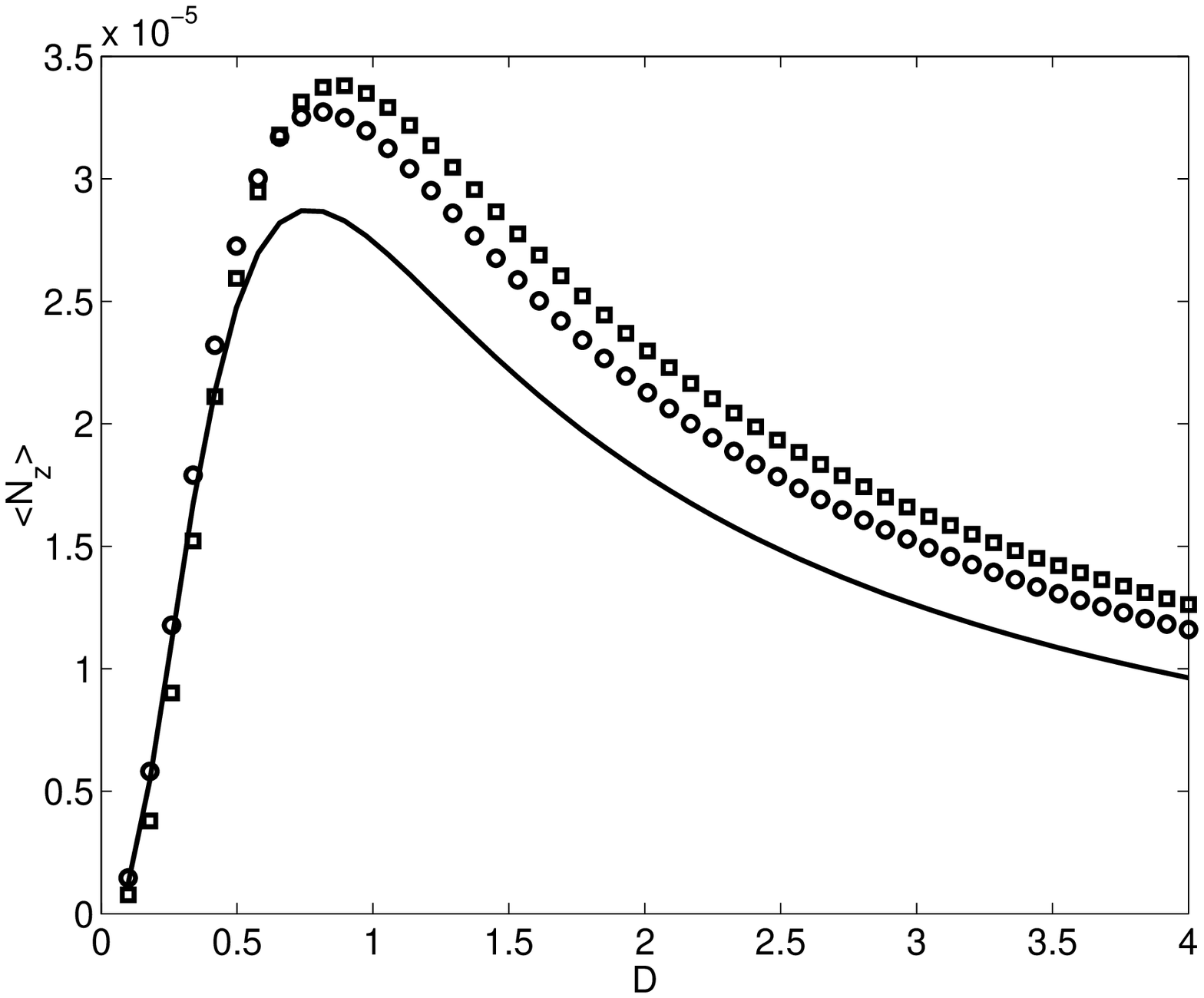}  
  \caption{\label{fig:perttest} Left: Time averaged torque $\Nav$ as a
    function of the noise intensity $D$ for the time dependence
    (\ref{foft1}) of the magnetic field. The parameter values are 
    $H_x=0.5 D,\, H_y^{(1)}=0.5 D,\, H_y^{(2)}=0.15 D$, and 
    $\delta=0$. The scaling of the magnetic field strength with $D$
    ensures that the perturbation parameter $\eps$ stays constant. Shown
    are the perturbative results (\ref{pertresN}) (circles) and
    (\ref{effresN}) (squares) together with the numerically exact
    result (full line). Right: Same for the time dependence
    (\ref{foft2}) of the oscillating field with 
    $H_y^{(0)}=H_y^{(1)}=0.5 D$. The approximate results are given by 
    (\ref{pertresN2}) and (\ref{effresN2}), respectively.} 
\end{figure} 

In the left panel of Fig.~\ref{fig:perttest} we compare the two
approximate expressions  (\ref{effresN}) and (\ref{pertresN}) for the
averaged torque with the 
results obtained from the numerical solution of the FPE. Displayed is 
the dependence of the torque on the noise intensity $D$ which is
different in the two approximations. In order to keep the expansion
parameter $\eps\sim H/D$ roughly constant when changing $D$ we have 
scaled the magnetic fields strengths with $D$. As can be seen, up to
values of $\eps\simeq 0.5$ the accuracy of (\ref{pertresN}) is rather
good whereas (\ref{effresN}) describes the torque only qualitatively. 

Proportional change of magnetic field strength and noise intensity is
equivalent to a change of the frequency $\om$ of the external 
field in (\ref{deffield}). Fig.~\ref{fig:perttest} hence also
demonstrates the resonance-like character of the investigated ratchet
effect. $\Nav$ is largest when the deterministic 
time scale of the external driving matches an intrinsic stochastic
time scale of the system related to the Brownian relaxation time
$\tau_B$  defined in (\ref{deftauB}). 

The two approximate expressions for the torque (\ref{effresN}) and
(\ref{pertresN}) differ in their dependence on the phase angle
$\delta$. When using (\ref{pertresN}) to fit the experimental results
reported in \cite{EMRJ} we find for the Brownian relaxation time the
value $\tau_B\simeq 6.4\; 10^{-4}$s instead of $\tau_B\simeq 1.8\;
10^{-3}$s as obtained on the basis of (\ref{effresN}) in \cite{EMRJ}.
The experiments were done with a ferrofluid with $\eta\simeq 0.1$ Pas. Using
(\ref{deftauB}) the two results for $\tau_B$ hence translate in fits for
the particle diameter of $d\simeq 26$nm and $d\simeq 36$nm,
respectively. These values exceed the typical diameter of roughly $10$
nm by factors between 2 and 3, respectively. The main reason for this
discrepancy is probably the polydispersedness of real ferrofluids,
having a particle size distribution with a long tail (see
{\it e.g.} \cite{Ros}, ch.2). Hence, a whole spectrum of relaxation times is
necessary to accurately describe the dynamics of the magnetization.
However, this issue is not at the focus of the present investigation. We also
note that the transferred torque typically increases with the particle
size and hence the described ratchet effect is likely to be dominated
by the larger grains in the ferrofluid. 

A similar approximate calculation of the average torque is also
possible for the time dependence (\ref{foft2}). The result of the
effective field approximation is 
\begin{equation}\label{effresN2}
  \Nav_e=\frac{2}{45}\,
     \frac{H_x\,H_y^{(0)}\,(H_y^{(1)})^2}{D\, (4D^2+1)^2}\ ,
\end{equation}
whereas we find from the perturbative solution of the FPE 
\begin{equation}\label{pertresN2}
  \Nav_p=\frac{H_x\,H_y^{(0)}\,(H_y^{(1)})^2}{30}\,
     \frac{44D^2+3}{D\, (4D^2+1)^2\,(36 D^2+1)}\ .
\end{equation}
The two expression only differ significantly from each other if the
noise intensity $D$ is very small. In the right panel of
Fig.~\ref{fig:perttest} they are compared with the result from the
numerical solution of the FPE (\ref{FPE}). For $\eps\simeq 0.5$ the
accuracy is again seen to be rather satisfactory


\section{Conclusions}\label{sec:conc}

In the present paper we have theoretically investigated the
rotational Brownian motion of colloidal ferromagnetic particles in
an oscillating magnetic field. The central tool
was the Fokker-Planck equation for the probability density $P(\e,t)$
of the particle orientation $\e$ at time $t$. Solving this equation
either numerically or approximately by using the effective field
approximation as well as a perturbative expansion we have determined
the time averaged torque $\Nav$ exerted by the magnetic field on the
particles. The main and a priori quite unexpected qualitative finding
is the fact that a purely oscillating magnetic field without net
rotating component can transfer angular momentum to a ferromagnetic
grain. As basic mechanism behind this transfer, a ratchet effect was
identified by which the magnetic field rectifies the thermal
fluctuations of the particle orientation that arise due to random
collisions with the molecules of the carrier liquid. The detailed
operation of this ratchet effect in the present system was discussed
on the basis of a weak-noise analysis of the Fokker-Planck equation
for a closely related one-dimensional system in section \ref{inst1d}. 

Via the viscous coupling to the carrier liquid the torque on the
particle is transmitted to the liquid. The combined {\em microscopic}
torques of the huge number of individual nanoparticles then yields a 
{\em macroscopic} rotation of the ferrofluid as a whole, as observed
experimentally in \cite{EMRJ}. In the absence of thermal noise, no 
average torque can arise, {\it i.e.} thermal fluctuations are an indispensable
requirement for the rotation of the individual grains and hence of the 
ferrofluid as a whole to occur. 

The results from the Fokker-Planck equation are completely
consistent with rather general symmetry considerations detailed in
section \ref{sec:symm}. Moreover, they agree very well with simulations
of the corresponding Langevin equations as given in \cite{RaEn} and
quantitatively describe the experimental findings reported in
\cite{EMRJ}. 

We remark that our present system puts forward a new type of thermal 
ratchet device which does not fit into any previously known 
specific class of ratchet systems. Adopting the classification scheme
from \cite{Reirev}, our present system has some similarity with
so-called asymmetrically tilting ratchets as well as with a so-called
traveling potential ratchets, however in the generalized sense
involving two counterpropagating traveling potentials. Yet there remain
significant differences with both these classes. In particular, we
note that we are not dealing here with a periodic, asymmetric
so-called ``ratchet''-potential. Rather, at any fixed instance of time the 
relevant potential is perfectly symmetric about the instantaneous
direction of the magnetic field and hence no preferential direction of
rotation seems to exists. It is only via the time evolution that a 
symmetry breaking arise, which is sometimes called a dynamical
symmetry breaking. 

In summary we hope that the present investigation has demonstrated
that ferrofluids are very suitable systems to study various aspects of
thermal ratchet behaviour, and that it may stimulate further
theoretical, numerical and experimental work in this direction.

\acknowledgements
Special thanks is due to Ubbo Felderhoff for several interesting
suggestions and to Martin Raible for providing 
the numerical simulation results in Fig. 3.
Discussions with Hanns Walter M\"uller,
Konstantin Morozov, and Mark Shliomis are gratefully
acknowledged. Part of this work was done during 
a sabbatical stay of A.E. at the Laboratoire de Physique Th{\'e}orique
at the Universit\'e Louis Pasteur in Strasbourg. He acknowledges the
kind hospitality as well as financial support from the 
{\em Volkswagenstiftung}. This work was further supported by Deutsche
Forschungsgemeinschaft under SFB 613 and RE 1344/3-1 and by the
ESF-program STOCHDYN.

\end{document}